\documentclass[letter]{bioinfo}
\copyrightyear{2011}
\pubyear{2011}

\usepackage{amsmath}
\usepackage{color}

\usepackage{graphicx}
\usepackage{amssymb}
\usepackage{multirow}

\newcommand{\eg}{{\it e.g.}}
\newcommand{\ie}{{\it i.e.}}
\newcommand{\argmax}{\operatornamewithlimits{arg\,max}}

\begin{document}
\firstpage{1}

\title[HiTRACE: High-throughput analysis for capillary electrophoresis]{HiTRACE: High-throughput robust analysis for capillary electrophoresis}
\author[Yoon \textit{et~al}]{Sungroh~Yoon\,$^{1*}$, Jinkyu~Kim\,$^{1}$, Justine~Hum\,$^{2}$, Hanjoo~Kim\,$^{1}$, Seunghyun~Park\,$^{1}$, Wipapat~Kladwang\,$^{2}$, and Rhiju~Das$^2$\footnote{to whom correspondence should be addressed}}
\address{$^{1}$School of Electrical Engineering, Korea University, Seoul 136-713, Republic of Korea\\
$^{2}$Departments of Biochemistry and Physics, Stanford University, Stanford, CA 94305, USA}

\history{}

\editor{}

\maketitle

\begin{abstract}

\section{Motivation:}
Capillary electrophoresis (CE) of nucleic acids is a workhorse technology underlying high-throughput genome analysis and large-scale chemical mapping for nucleic acid structural inference. Despite the wide availability of CE-based instruments, there remain challenges in leveraging their full power for quantitative analysis of RNA and DNA structure, thermodynamics, and kinetics. In particular, the slow rate and poor automation of available analysis tools have bottlenecked a new generation of studies involving hundreds of CE profiles per experiment.

\section{Results:}
We propose a computational method called \emph{high-throughput robust analysis for capillary electrophoresis} (HiTRACE) to automate the key tasks in large-scale nucleic acid CE analysis, including the profile alignment that has heretofore been a rate-limiting step {\color{red} in the highest throughput experiments.  We illustrate the application of HiTRACE on thirteen data sets representing 4 different RNAs, three chemical modification strategies, and up to 480 single mutant variants; the largest data sets each include 87,360 bands.} By applying a series of robust dynamic programming algorithms, HiTRACE outperforms prior tools in terms of alignment and fitting quality, as assessed by measures including the correlation between quantified band intensities between replicate data sets. Furthermore, while the smallest of these data sets required 7 to 10 hours of manual intervention using prior approaches, HiTRACE quantitation {\color{red} of even the largest data sets herein was achieved in 3 to 12 minutes}. The HiTRACE method therefore resolves a critical barrier to the efficient and accurate analysis of nucleic acid structure in experiments involving tens of thousands of electrophoretic bands.

\section{Availability:}
HiTRACE is freely available for download at http://hitrace.stanford.edu.
\section{Contact:} \href{sryoon@korea.ac.kr, rhiju@stanford.edu}{sryoon@korea.ac.kr, rhiju@stanford.edu}
\end{abstract}

%

\section{Introduction}\label{s:introduction}
\begin{figure*}
\centering
\includegraphics[width=0.9\linewidth]{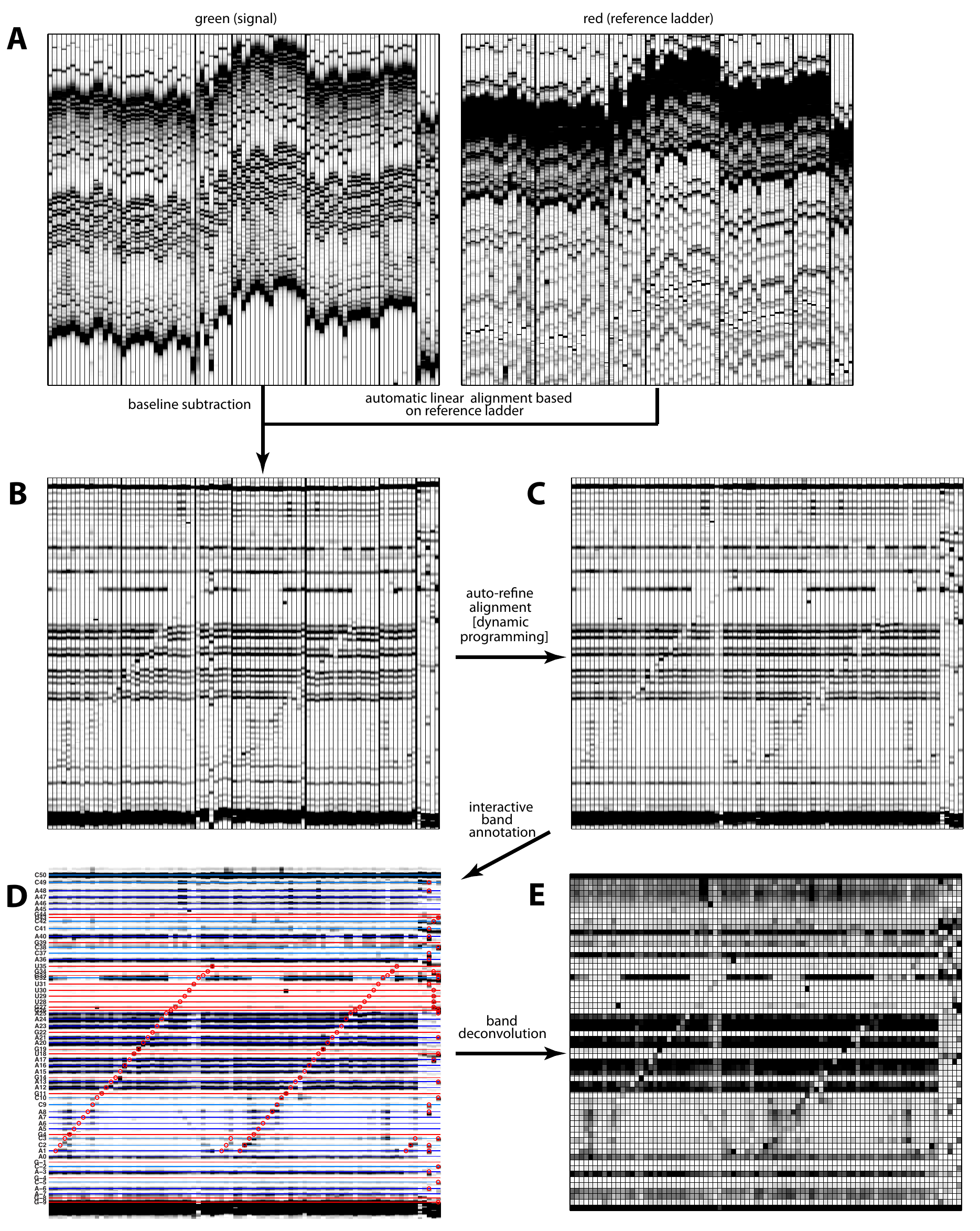}
\caption{Overview of the proposed HiTRACE methodology. { \color{red} (A) Raw electropherograms for an example data set. (left) Dimethyl sulfate (DMS) modification of the MedLoop RNA~\citep{kladwangcordero2011}, read out by reverse transcription with rhodamine-green-labeled primers followed by DNA separation by capillary electrophoresis; data shown are DMS profiles for 80 (of 120) single nucleotide mutants, two replicate controls without chemical modification, and sequencing ladders for C, U, and G. (right) Electropherograms of the Texas-red-labeled DNA ladder that was co-loaded with each sample to produce fiducial markers for alignment.  (B) Profiles after automated preprocessing (baseline subtraction) and correlation-optimized linear alignment. (C) Profiles after automated alignment refinement by dynamic-programming-based nonlinear adjustments. (D) Interactive sequence annotation guided by features (red circles) at mutation positions and bands in the sequencing ladder. Blue, cyan, orange, and red lines correspond to modifications at A, C, G, and U, respectively. (E) Quantitated band areas (the final output of HiTRACE). Shorter DNA fragments (higher mobility) are at the top of each panel.} }
\label{f:workflow}
\end{figure*}

Capillary electrophoresis (CE) is a widely used approach for biochemical analysis. The rapid electrophoretic separation of fluorescently labeled nucleic acid fragments inside electrolyte-filled capillaries significantly accelerated genome sequencing~\citep{ruiz1993dna,woolley1995ultra}. A more recent, powerful application of CE enables the high-throughput structure analysis of self-assembling nucleic-acid-containing systems~\citep{mitra2008high,vasa2008shapefinder,weeks2010,daskaranicolasbaker2010,kladwang2010} as complex as viruses~\citep{weeksnature2009,weeksplos2009} and ribosomes~\citep{deigan2009accurate} at single-nucleotide resolution.

The CE profiles obtained in this recent generation of `structure-mapping' experiments present tens of thousands of individual electrophoretic bands; quantifying these data gives detailed portraits of nucleic acid structure, folding thermodynamics, and kinetics but requires significant informatics efforts~\citep{mitra2008high}. `Base-calling' software packages can assign sequences to these bands in special four-color experiments (see, \eg, ~\citealp{phred1,phred2}) but are not applicable to structure mapping experiments, which require more robust sequence annotation and quantitative fits of each profile to a sum of peak shapes. {\color{red}Such quantitative analysis} is aided by the design of experiments so that the desired information appears as differences between corresponding bands across profiles [see, e.g., ~\citep{das2005safa,kladwang2010}]; then, sequence annotation of one profile results in annotation of corresponding bands across the entire data. For these data sets, tools for \emph{alignment of features}, or `rectification'~\citep{das2005safa,laederach2008semiautomated}, across different profiles resulted in improvements in quantification speed and accuracy, but these tools remain poorly automated. As the experimental steps of large-scale CE measurements continue to accelerate, the bioinformatic task of profile alignment has become a rate-limiting step in carrying out these information-rich structural studies.

Current approaches to aligning and fitting capillary profiles include capillary automated footprinting analysis (CAFA; \citealp{mitra2008high}) and ShapeFinder~\citep{vasa2008shapefinder}; we have found these methods difficult to apply to large-scale titration or mutate-and-map data sets~\citep{kladwang2010,kladwangcordero2011}. For instance, CAFA is focused more on peak fitting and has limited alignment capabilities. The ShapeFinder alignment function can align spectrally separated products within a single capillary but not profiles across multiple capillaries with initially poor alignment. Use of these tools requires tedious manual intervention and risks bias or unnecessary errors from such manipulation. Analysis tools for alignment and peak fitting have also been proposed in other domains such as chromatography~\citep{nielsen1998aligning,tomasi2004correlation}, mass spectrometry~\citep{wong2005specalign,kazmi2006alignment} and slab gel electrophoresis~\citep{das2005safa,laederach2008semiautomated}, but, empirically, these approaches give unsatisfactory performance for CE data.

To address the limitations of existing methods, we have developed \emph{high-throughput robust analysis for capillary electrophoresis} (HiTRACE) to automate the alignment and quantification of nucleic acid structure mapping profiles obtained from hundreds of capillaries. As depicted in Figure~\ref{f:workflow}, the proposed method consists of four major steps: preprocessing (step A), correlation-optimized linear alignment (step B), dynamic-programming-based nonlinear adjustments (step C), sequence annotation (step D), and peak fitting (step E). After describing the core algorithms that underlie the robust automation of each step, we present quantitative comparisons illustrating the substantial boosts in both accuracy and speed of HiTRACE over previous approaches. With the proposed methodology, the previously rate-limiting step of quantifying high-throughput CE data is now faster than experimental data acquisition times, enabling the investigation of nucleic acid structure at an unprecedented rate.

\begin{methods}
\section{Methods}\label{s:method}
\subsection{Experimental setup}

{\color{red} An experimental protocol that is optimal for HiTRACE alignment and quantification has been developed; complete descriptions of reaction components, purification procedures, and sequencing ladder generation have been given previously \citep{daskaranicolasbaker2010,kladwang2010,kladwangcordero2011}. Briefly, RNA samples were chemically modified under the desired solution conditions and then reverse transcribed with primers (labeled at the 5$^{\prime}$ ends with the rhodamine green fluorophore) complementary to the 3$^{\prime}$ end of the RNA. Because the reverse transcription stops at modified nucleotides, the length distribution of the resulting DNA products encodes the chemical reactivities of the RNA. Length separation of the DNA was carried out on Applied Biosystems ABI 3100 and ABI 3730 sequencers; these intruments permit the single-nucleotide separation of products as long as 500 nucleotides for 16 and 96 samples, respectively. To facilitate HiTRACE alignment, all samples were co-loaded with a reference ladder that fluoresces in a different color and provides fiducial markers that are identical between samples. The ladder, prepared in a large batch for many experiments, was derived by reverse transcribing an arbitrary RNA (typically the 202-nucleotide P4-P6 RNA) with a Texas-red-labeled primer.}

\subsection{Assumptions and definitions}
CE profiles each contain hundreds of `bands' (when the data are viewed in gray scale) or `peaks' (when the intensity is plotted as a function of electrophoresis time) whose intensities or areas report on individual residues of a nucleic acid sequence. In CE experiments that use hundreds of capillaries, profiles are typically obtained in multiple batches of experiments, \eg, with 16 capillaries in an ABI 3100 sequencer, as illustrated in Figure~\ref{f:workflow}. The first profile of each batch is designated the reference to which other profiles of the batch should be aligned.
Each profile $i$ represents fluorescence intensity measured at uniformly spaced time points (here, 0.1 seconds) denoted by $n = 1, 2, \ldots, N$ with associated intensity values $y_i(n)$. As shown in Figure~\ref{f:workflow}, the horizontal and vertical axes correspond to the profile index and the measurement position in timepoints, respectively. Fluorescence intensity levels are represented in gray scale, with nucleic acid species of different lengths appearing as separated, dark bands.
The desired final output of the proposed methodology is a set of aligned profiles with their quantified band areas.

\subsection{Preprocessing (step A)}
In a typical profile, the starting and ending regions contain no signal. To accelerate subsequent steps, the user has the option of defining a window that brackets the electrophoretic signals in all the profiles. As another preprocessing step, we subtract an offset, constant within each profile, so as to bring the signal to zero at the boundaries of the window; this step corrects for overall drift in signal baselines that are observed in sequencer detectors. We have also implemented an option to derive and subtract a smooth (but not necessarily linear) baseline from each profile by using a procedure similar to \citet{xi2008baseline}. This operation removes smoothly varying backgrounds in fluorescence signal sporadically seen in experimental CE profiles and, empirically, brings independent replicates into closer agreement.

\subsection{Alignment by linear transformation (step B.1)}
The first step involves a linear scaling and shifting of the time axis based on maximizing the correlation coefficient between each fluorescence profile $y_i(n)$ and the reference profile $y_1(n)$ within each batch: \begin{equation}
(\Delta_i^*, \sigma_i^*) = \!\!\!\!\argmax_{(\Delta_i,\sigma_i) \in D_i\times S_i} \left\{\mathrm{corr}\left[y_1(n), y_i\left(\frac{n}{\sigma_i}-\Delta_i\right)\right]\right\}\label{e:align_corr}
\end{equation}
where $D_i$ and $S_i$ represent the sets of possible values of the shift $\Delta_i$ and scale factor $\sigma_i$, respectively, and $D_i\times S_i$ denotes their Cartesian product. Based on the values found above, we first time-scale each profile $y_i(n)$ by $\sigma_i^*$ using linear interpolation and then shift it by $\Delta_i^*$.
The correlation coefficient was chosen as the optimization target because it is independent of signal offset and scaling and has been widely used in other alignment tasks \citep{nielsen1998aligning,bylund2002chromatographic,pravdova2002comparison,tomasi2004correlation}. 
We carry out the search over shifts ($\Delta_i$) efficiently through a Fast Fourier Transform (FFT; \citealp{oppenheim09}). By default, we carry out the alignment based on the reference ladder pattern that is co-loaded with each sample (see above).

\subsection{Alignment between batches (step B.2)}
 Due to variabilities between batches, performing only the intra-batch alignment above produces stratified alignment results, where a number of up-and-down `stairs' appear. To resolve this problem, we perform an additional \emph{inter-batch} alignment.
 This step constructs a representative profile of each batch by calculating the average of the first, middle, and last profiles selected from each batch. We align these representative profiles to the first representative profile by the procedure above. Assume that, for the representative profile from batch $b$, we have determined $\Delta^*_b$ and $\sigma^*_b$ values. We then re-align all the profiles in batch $b$ using these $\Delta^*_b$ and $\sigma^*_b$ values (see Figure~\ref{f:workflow}B). More details of step B.2 can be found in the supplement.

\subsection{Nonlinear alignment (step C)}
With current CE equipment, we found that it was not feasible to get complete alignment for profiles with single-band resolution through just linear scaling of the time axis. There are two reasons for this problem. First, the electrophoretic mobilities of the same products in different capillaries, or for the same capillary used at different times, can vary due to temperature differences and geometry differences. As a result, long profiles, containing hundreds of bands measured over tens of minutes, can be aligned well over the initial part of the data (\eg, the first two minutes)  or the final part of the data (\eg, the last two minutes), but both parts cannot be simultaneously aligned with a single linear transformation. Second, we often run capillary electrophoresis experiments for structure mapping of molecules with slightly different sequences, \eg, libraries of single-mutation constructs~\citep{kladwang2010,kladwangcordero2011}. This leads to small perturbations in the band mobilities at the site of the mutation and requires a locally nonlinear transformation to permit alignment. To correct for both these issues, we perform another round of refining the alignment.

The concept underlying the non-linear alignment is depicted in the supplement, and resembles the warping method presented in \citet{nielsen1998aligning} for chromatographic data. We break the time axis of a non-reference profile into $m$-pixel windows and then shift each window boundary within a predefined range over the reference profile so as to maximize the correlation between profiles summed over all windows. We assume that the window ordering is preserved during alignment. The number of possible arrangements in this setup is large but can be enumerated efficiently by a dynamic programming (DP) approach~\citep{cormen2009introduction,nielsen1998aligning,bylund2002chromatographic,robinson2007dynamic} that recursively solves the problem for the first window, then the first two windows, etc. As in steps B.1~\&~B.2, we accelerate the calculation by computing correlation coefficients through FFT. Example results are shown in Figure~\ref{f:workflow}C. The supplement includes a graphical description of determining the shift amount for each window edge for aligning two example profiles.

\begin{table}
\processtable{High-throughput RNA structure mapping data sets analyzed by HiTRACE.\label{t:data}}
{\begin{tabular}{llcccc}
\toprule
Name& \# profiles & \# bands per profile & \# total bands \\
\midrule
X20/H20 DMS-1$^a$  &  98 & 40  & 3920 \\
X20/H20 DMS-2$^a$  &  88 & 40  & 3520 \\
MedLoop DMS-1$^b$  & 120 & 60  & 7200 \\
MedLoop DMS-2$^b$  & 136 & 60  & 8160 \\
MedLoop CMCT-1$^b$ & 128 & 60  & 7680 \\
MedLoop CMCT-2$^b$ & 120 & 60  & 7200 \\
SRP DMS-1$^c$      &  88 & 60  & 5280 \\
SRP DMS-2$^c$      &  96 & 60  & 5760 \\
SRP CMCT-1$^c$     &  88 & 60  & 5280 \\
SRP CMCT-2$^c$     &  88 & 60  & 5280 \\
P4-P6 DMS$^c$      & 480 & 182 & 87360\\
P4-P6 CMCT$^c$     & 480 & 182 & 87360\\
P4-P6 SHAPE$^c$    & 480 & 182 & 87360\\
\botrule
\end{tabular}}
{$^a$\citealp{kladwang2010}; $^b$\citealp{kladwangcordero2011}; $^c$This work.\\
Abbreviations: SRP, signal recognition particle conserved domain; P4-P6, P4-P6 domain of the Tetrahymena group I ribozyme; DMS, dimethyl sulfate; CMCT, 1-cyclohexyl-3-(2-morpholinoethyl) carbodiimide metho-p-toluenesulfonate; SHAPE, selective hydroxyl acylation analyzed by primer extension.}
\end{table}

\subsection{Sequence annotation (step D)}
Each band in a fluorescence profile corresponds to a position in the nucleic acid sequence. Currently, we carry out sequence annotation interactively and manually, as this encourages visual inspection of the data and makes use of expert knowledge to ensure accurate annotation.{  \color{red} This step is accelerated compared to prior approaches \citep{mitra2008high,vasa2008shapefinder} through visual feedback. Sequence assignments are made based on Sanger sequencing ladders included in the experiments; as the user makes assignments, `guidemarks' appear at expected band positions (one residue longer than the corresponding position of modification, due to dideoxynucleotide incorporation). These guidemark positions can be visually confirmed or adjusted to overlay on experimental bands  (see circles in Fig.~\ref{f:workflow}D). In addition, these guidemarks can be set to appear on A and C positions for dimethyl sulfate alkylation experiments~\citep{peattie1980chemical,tijerina2007dms}, as well as mutated positions in mutate-and-map experiments (Fig.~\ref{f:workflow}D), which typically give visually distinct perturbations in chemical modification. These features provide cross-checks on the sequencing ladder that confirm accuracy. Due to the alignment of traces achieved in previous steps, sequence annotations need to only be provided once and are applicable to all traces. Automated annotation procedures are also being developed and will be incorporated in future versions of HiTRACE. }

\subsection{Band deconvolution and quantification (step E)}
In this last step of HiTRACE, we approximate each profile $y(n)$ for $n=1,2,\ldots,N$ by a sum $f(n)$ of $K$ Gaussian curves with the form
\begin{equation}
f(n) = \sum_{k=1}^K A_k \exp\left[-\frac{(n-\mu_k)^2}{2\sigma_k^2}\right]
\end{equation}
such that the deviation defined by
\begin{equation}
\sqrt{\frac{1}{N}\sum_{n=1}^N\left[f(n) - y(n) \right]^2}
\end{equation}
is minimized. $A_k$, $\mu_k$ and $\sigma_k$ are the parameters that determine the amplitude, the center location and the width, respectively, of a peak modeled by a Gaussian. We find the optimal values of these parameters by a standard Levenberg-Marquardt optimization technique for least-square minimization~\citep{levenberg1944method,marquardt1963algorithm}, and report the area of each peak as the final output.

\subsection{Implementation and data preparation}
We implemented the proposed HiTRACE methodology in the MATLAB programming environment (The MathWorks, http://www.mathworks.com) and are making it freely available for download at http://hitrace.stanford.edu. For comparison with HiTRACE, we also prepared the implementations of the five different profile analysis algorithms: CAFA~\citep{mitra2008high}, ShapeFinder~\citep{vasa2008shapefinder}, msalign~\citep{kazmi2006alignment}, SpecAlign~\citep{wong2005specalign}, and COW~\citep{tomasi2004correlation}. We could not apply some methods to all situations due to their intrinsic limitations. For instance, the alignment feature of CAFA and ShapeFinder requires significant manual intervention to handle hundreds of profiles; we did not include ShapeFinder in the alignment result comparison. Similarly, msalign, SpecAlign, and COW can align profiles but do not carry out peak fitting. We thus excluded them in fitting result comparisons.

\subsection{Criteria for evaluating alignment results}\label{ss:criteria}
We applied two widely used mathematical criteria---the mean squared error (MSE;~\citealp{kay1993fundamentals}) of aligned peak positions with respect to the reference peaks and the Kullback-Leibler (KL) divergence~\citep{cover2006elements} between reference and non-reference profiles.

In MSE computation, we consider the position $p$ of each peak in the reference profile as the true value being estimated, and use the position $\hat{p}$ of the aligned peak on a non-reference profile as the estimator of $p$. The MSE for the $j$-th reference peak $p_j$ is then
\begin{equation}
MSE_{j} = E\left[(\hat{p}_j - p_j)^2\right] = \frac{1}{L}\sum_{i=1}^L\left(\hat{p}_{ij}-p_{1j}\right)^2
\end{equation}
where $L$ is the number of profiles in the data set used, and $p_{1j}$ and $\hat{p}_{ij}$ represent the positions of the $j$-th reference peak and the peak on profile $i$ that is aligned to $p_j$, respectively. For the peak detection step involved in the MSE computation, we used the peak algorithm described by~\citet{mitra2008high}, which is specifically designed for finding peaks in CE profiles and shows satisfactory performance for our purpose.

To evaluate the alignment results from an information-theoretic perspective, without explicitly considering specific peaks or band positions, we utilized the KL divergence.
We calculated the KL divergence between the reference profile $y_1(n)$ and a non-reference profile $y_i(n)$ as
\begin{equation}
D_{KL}(y_1||y_i) = \sum_{n=1}^N y_1(n) \log \frac{y_1(n)}{y_i(n)}
\end{equation}
where $N$ is the number of pixels in each profile. We repeat this calculation for every reference and non-reference pair in a data set. Before computing KL divergence, intensity values were limited to two standard deviations above the mean to prevent KL divergence values from being dominated by strong bands at the beginning and end of each profile.

\end{methods}

\section{Results}\label{s:result}
{\color{red} \subsection{High-throughput RNA structure mapping data sets}
To test HiTRACE, we collected 13 nucleic acid structure mapping experiments read out by capillary electrophoresis (Table~\ref{t:data}). These data sets were diverse: probed molecules included artificial model systems (the MedLoop RNA and the X20/H20 RNA/DNA system) as well as natural structured RNAs (a conserved domain from the signal recognition particle and the P4-P6 domain of the Tetrahymena group I ribozyme), with lengths between 60 and 202 nucleotides. Three common chemical modification strategies were represented in the data: dimethyl sulfate alkylation \citep{tijerina2007dms}, carbodiimide modification \citep{walczak}, and $2^{\prime}$-OH acylation [the SHAPE strategy \citep{wilkinson2005}]. In addition, the data sets were challenging in their size. Three experiments each gave 182 bands over 480 electropherograms, for a total of 87,360 bands per data set. Finally, to test the precision of quantification relative to other sources of error, five experiments were conducted twice by two independent researchers. Additional data sets were collected to confirm HiTRACE's ability to quantify data for RNAs over 400 nucleotides in length (the L-21 ScaI Tetrahymena group I ribozyme) and to compare overlapping SHAPE data derived from reverse transcription starting at different primers on the same RNA (the P4-P6 domain). Overall, these data sets provide a diverse and challenging benchmark of nucleic acid CE experiments at the large scale permitted by current high-throughput experimental protocols. }

\begin{figure}
\centering
\includegraphics[width=\linewidth]{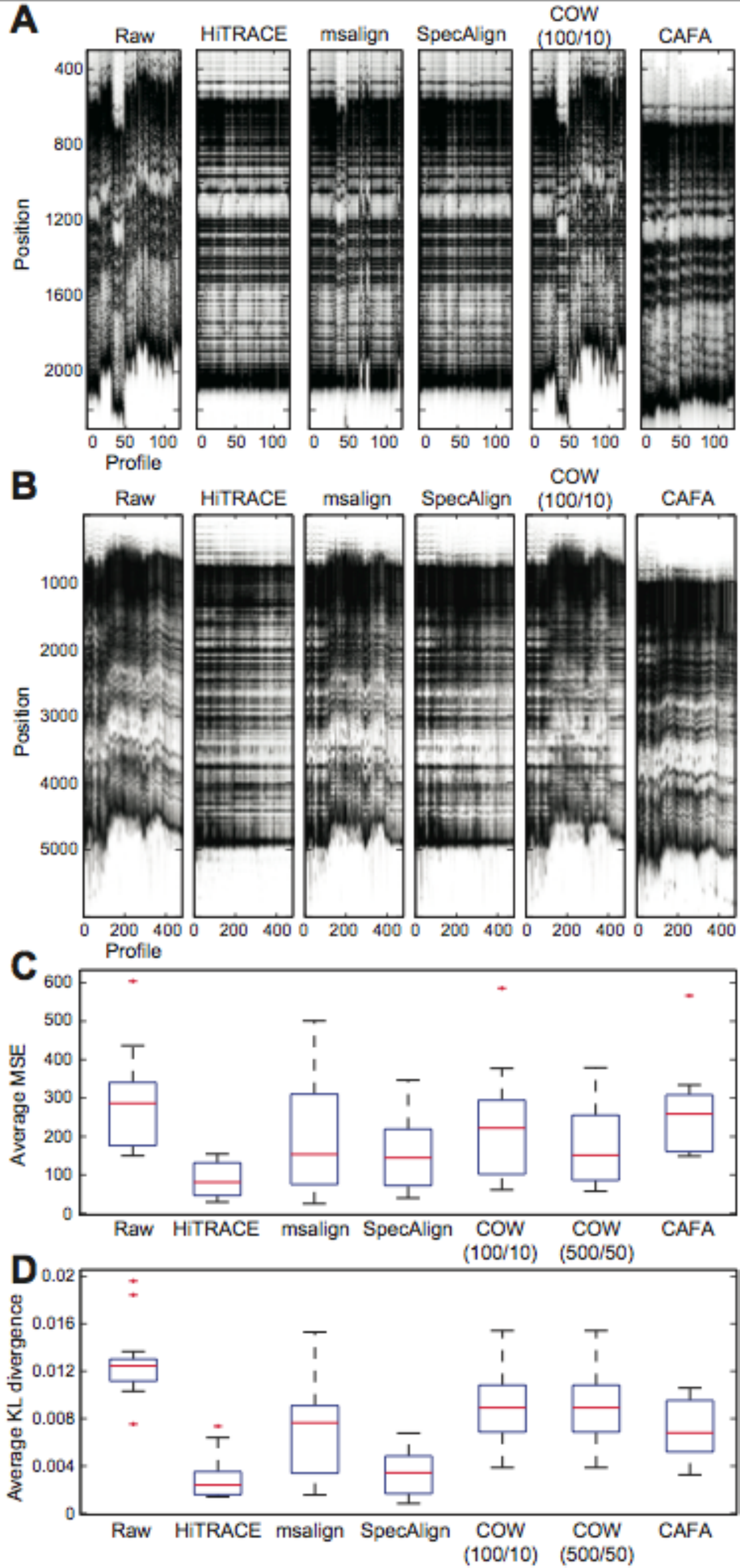}
\caption{ Comparison of available alignment strategies for nucleic acid CE profiles.  (A) Comparison of electrophoretic profiles of the 88-profile MedLoop DMS mutate-and-map data set~\citep{kladwangcordero2011} (replicate 2) before alignment and after alignment by HiTRACE, msalign~\citep{kazmi2006alignment}, SpecAlign~\citep{wong2005specalign}, COW~\citep{tomasi2004correlation} and CAFA~\citep{mitra2008high}. (B) Alignment results for the 480-profile P4-P6 DMS data set.  {\color{red}(C) Quantitative comparison of alignment results for all 13 data sets based on mean squared error (MSE;~\citealp{kay1993fundamentals}) of aligned peak positions with respect to the reference peaks. Red line is median value; box boundaries represent 75th and 25th percentiles; error bars represent the most extreme values whose distance from the box is less than 1.5 times the box length; + symbols are outliers beyond this range. (D) Quantitative comparison in terms of the Kullback-Leibler divergence~\citep{cover2006elements} between reference and non-reference profiles.} }
\label{f:alignment}
\end{figure}

\subsection{Robust alignment of CE profiles}
As the most basic test, we first compared the alignment results of HiTRACE with previously available methodologies by visual inspection (Figure~\ref{f:alignment}). Prior to alignment, CE experiments gave initially poor alignments of DMS chemical mapping profiles for the 60-band MedLoop RNA and the 182-band P4-P6 RNA  (``Raw'' in Fig.~\ref{f:alignment}A \& B). Application of automated HiTRACE alignment aligns the strong bands across all profiles (``HiTRACE'' in Fig.~\ref{f:alignment}A \& B; see also Fig.~\ref{f:workflow}A-C). In the alignment results produced by methods other than HiTRACE, profiles within each group tend to be reasonably aligned whereas profile groups are not well-aligned. We did not observe this `stratification' problem in the HiTRACE result, mainly due to the inter-batch alignment step (B.2) used by HiTRACE. Additionally, comparing HiTRACE results with SpecAlign and CAFA results reveals the effectiveness of the HiTRACE nonlinear alignment step, which adapts alignment to weakly varying electrophoretic rates along the profile. In the alignment results produced by SpecAlign and CAFA, some parts of the profiles appear reasonably aligned, but the top (SpecAlign) or bottom (CAFA) portions are not well-aligned. For msalign and COW, this problem is much more noticeable.

For more quantitative evaluation of profile alignments, we compared the different methodologies in terms of two mathematical criteria, mean squared error in peak position (MSE) and KL divergence between profiles. Figures~\ref{f:alignment}C \& D show the distributions of the average MSE and KL divergence values over the 13 data sets used for different algorithms. With respect to HiTRACE, the alternative methodologies produced poorer results, 1.73--3.09 and 1.51--3.94 times higher median MSE and KL divergence values, respectively.

\subsection{Leveraging accurate alignments into accurate quantification}
\begin{figure}
\centering
\includegraphics[width=\linewidth]{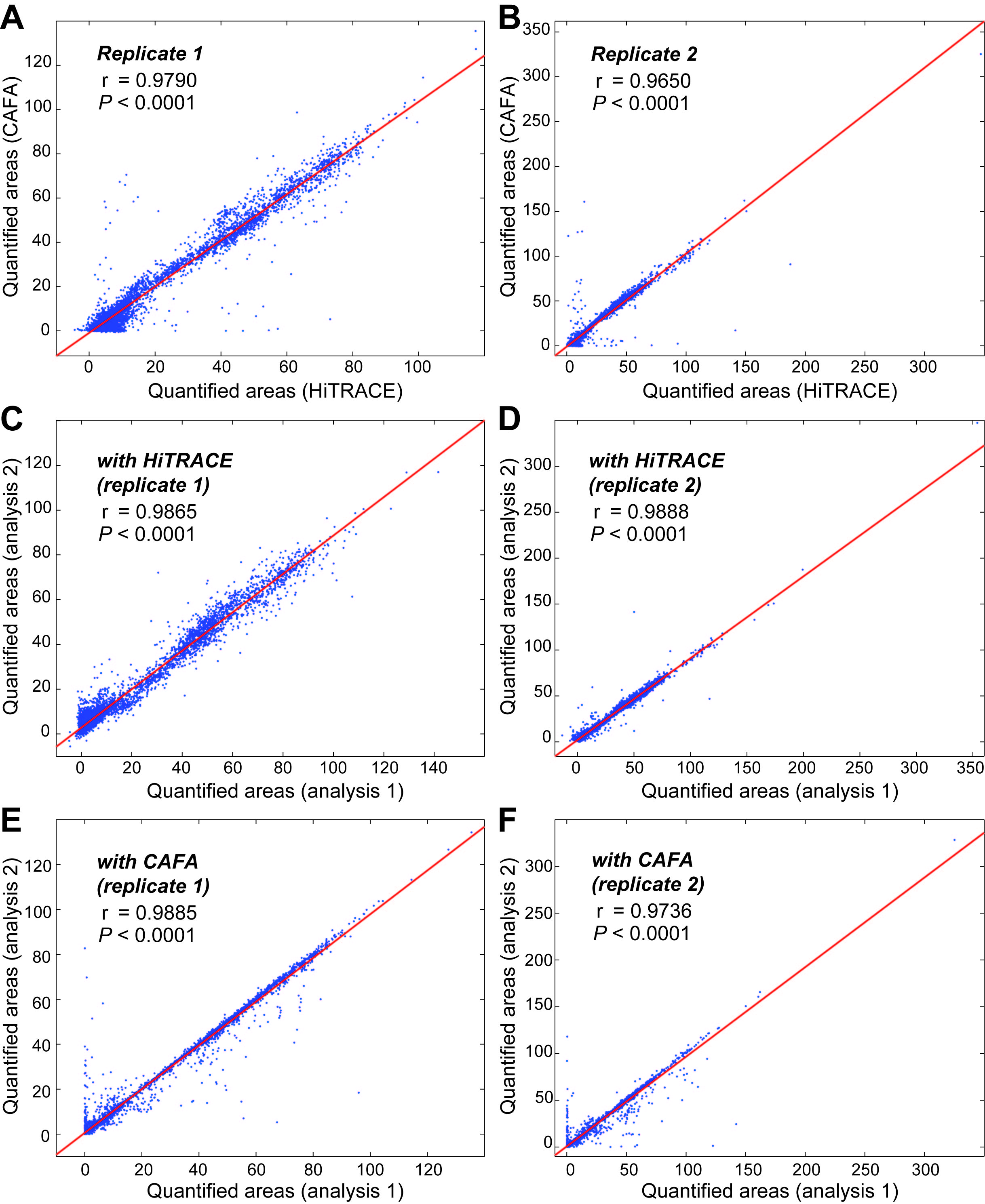}
\caption{ \color{red}Quantification accuracy and precision for HiTRACE and CAFA. Correlation of HiTRACE and CAFA results for two MedLoop DMS data sets [(A)~\&~(B)] confirms the absence of any systematic deviation between the two approaches. Precision of HiTRACE [(C)~\&~(D)] is similar or better than CAFA [(E)~\&~(F)], based on independent analyses of the same data set. }
\label{f:HiTRACE_CAFA}
\end{figure}

{\color{red} To assess the accuracy of the entire quantification procedure, including alignment, sequence annotation, and band deconvolution, we compared final quantified results between HiTRACE and previously available software for RNA structure mapping CE data, using two MedLoop DMS mutate-and-map data sets~\citep{kladwangcordero2011} (see also supplement for a comparison with the X20/H20 DMS data). Each set contained at least 120 profiles with 60 bands, for a total of 7200 data points per set. (Further comparisons between software packages were precluded by the difficulty of carrying out the analysis with prior software: ShapeFinder gave poor alignment even after several hours of manual intervention, and CAFA analysis required 10 hours of manual adjustment.)

The MedLoop sets gave excellent Pearson correlation coefficients between band intensities quantified with HiTRACE to those quantified with CAFA ($r$ of 0.979 and 0.965; Fig.~\ref{f:HiTRACE_CAFA}A\&B), confirming the lack of any major systematic errors introduced by the HiTRACE method. We hypothesized that the small, residual variance between the methods might stem from user-introduced variation during alignment (CAFA) or sequence assignment of bands (in CAFA and HiTRACE). To test this hypothesis, we carried out replicate quantification of the same data sets; the second independent analysis gave values with correlation coefficient ($r$) to the first analysis of 0.987 and 0.989 (HiTRACE; Fig.~\ref{f:HiTRACE_CAFA}C~\&~D) and 0.989  and 0.974 (CAFA; Fig.~\ref{f:HiTRACE_CAFA}E~\&~F). We conclude that any differences between HiTRACE and CAFA can be explained by imprecision (variance of 1.1--1.3\% in HiTRACE and 1.1--2.6\% in CAFA) introduced by users; this error is much smaller than variances arising from experimental error, as is discussed next.
}

\vspace{-0.25cm}

\subsection{Consistency in band quantification between experimental replicates}

\begin{figure}
\centering
    \includegraphics[width=\linewidth]{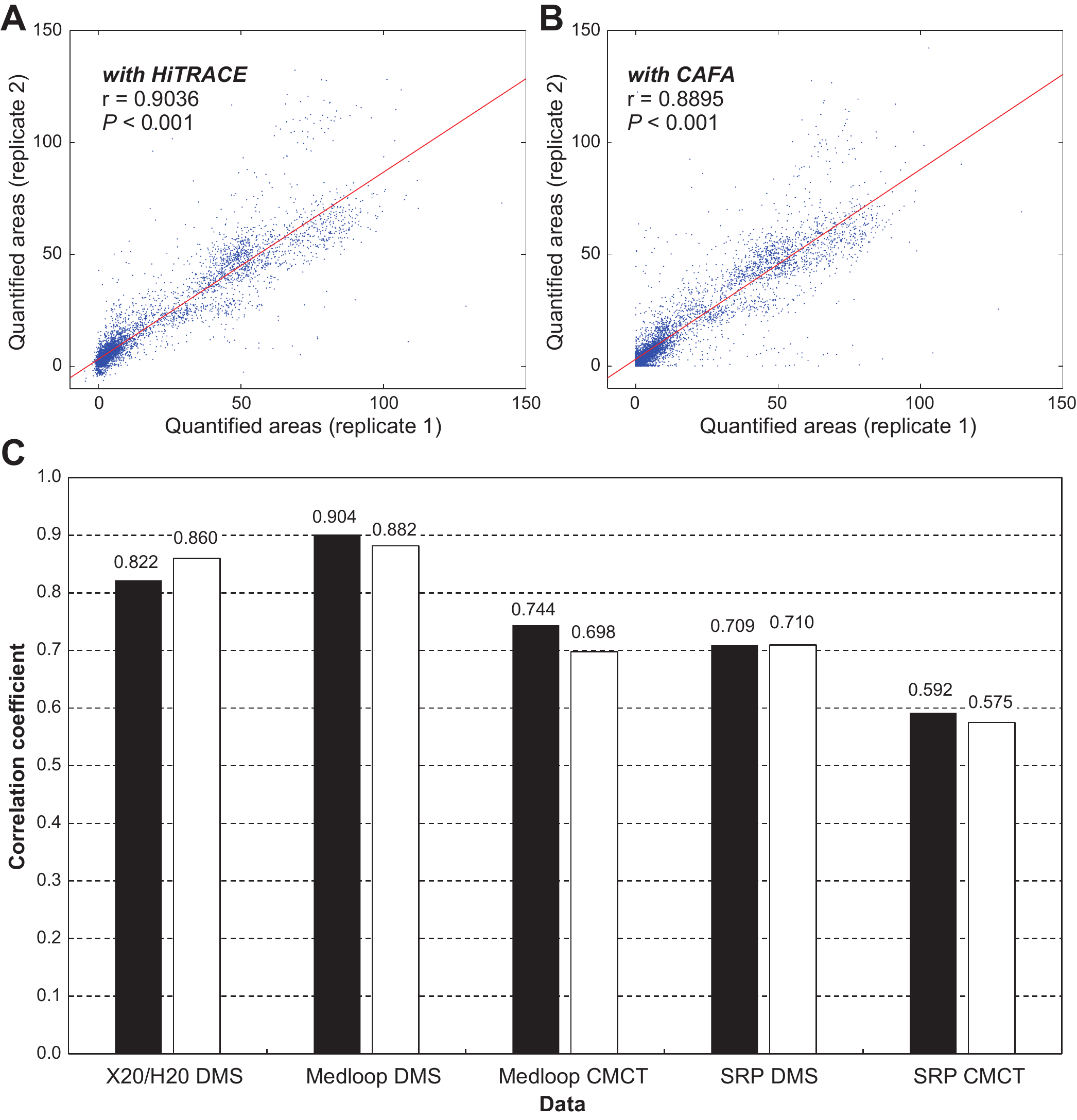}
\caption{\textcolor{red}{Correlation of results between experimental replicates for HiTRACE (A) and CAFA (B) for the MedLoop DMS mutate-and-map experiments~\citep{kladwangcordero2011}, and (C) for HiTRACE on five replicate data sets without (black bars) and with (white bars) optimization of Gaussian positions during band deconvolution.}}
\label{f:replicate-quant}
\end{figure}

{ \color{red} A stringent measure of the accuracy of an experiment and its analysis is the correlation of quantified intensities between independent replicates.} The goodness of this correlation is determined by experimental factors, including small variations in sample purity, pipetting errors, temperature differences, and variable times of each experimental step, and is also sensitive to any uncertainties arising from the data analysis procedure. {\color{red} We compared correlation coefficients between separate independent replicates of the MedLoop DMS mutate-and-map experiments~\citep{kladwangcordero2011}, quantified by both HiTRACE and CAFA (Fig.~\ref{f:replicate-quant}A\&B). In both cases, the cross-replicate correlations (0.89--0.90) are significantly lower than the intra-replicate comparisons (0.97--0.99) above, verifying that variances in experimental procedures exceed any variances in the data quantification.

The throughput of HiTRACE quantification enabled us to carry out this cross-replicate comparison for the additional replicate sets (see supplement) and to explore whether alternative data processing schemes might improve the precision of the HiTRACE quantification. We tested a computationally expensive band deconvolution procedure [previously used in SAFA~\citep{das2005safa}] that optimized centers of fitted Gaussians for each individual profile.  We observed indistinguishable cross-replicate correlation coefficients (Fig.~\ref{f:replicate-quant}C) with this procedure as compared to the the default HiTRACE method (no peak refinement). This comparison further validated the high quality of the profile-to-profile alignment in earlier HiTRACE steps, and motivated our choice to make as the HiTRACE default the 10- to 100-fold faster band-deconvolution procedure without band position fitting. We observed similarly invariant or slightly worse correlation coefficients in experiments without the baseline subtraction procedure; with additional alignment steps of `binarized' profiles; and with other methods to automatically refine band positions in each profile (see supplement).
}

\subsection{Reduced time demand of quantification}

\begin{figure}
\centering
\includegraphics[width=\linewidth]{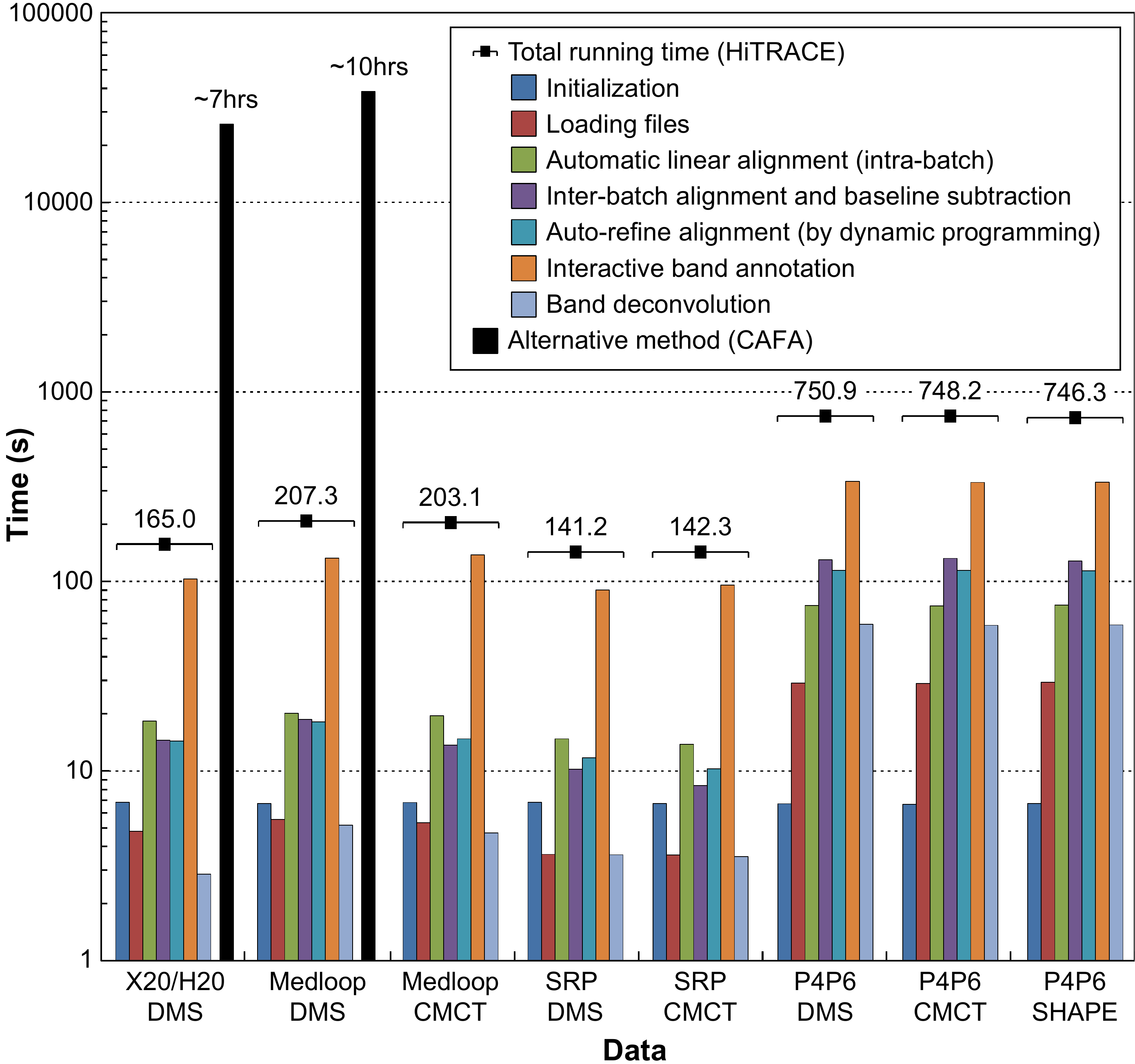}
\caption{The average running time of HiTRACE on different data sets. The time was measured on a personal computer system equipped with a \textcolor{red}{2.66GHz Core i5 processor (4 cores; multithreading enabled)} with 4GB RAM.
}
\label{f:runtime}
\end{figure}

Although HiTRACE relies on multiple steps for accurate analysis, the time demand of quantification by HiTRACE was considerably smaller (a few minutes) than the time required by prior informatic approaches as well as the time involved in preparing and obtaining the CE experiments (a few hours). Figure~\ref{f:runtime} shows the average running time of HiTRACE for different data sets, along with the breakdown of the running time. The largest data set (P4-P6 CMCT; 480 profiles {\color{red} and 87360 bands}) took approximately {\color{red}twelve minutes} to quantify, and the smaller sets (88--136 profiles, \textcolor{red}{4000--8000 bands) required three minutes or less}. Overall, HiTRACE averaged \textcolor{red}{1.58} seconds per profile from beginning (raw data load-in)  to end (quantified band intensities). For the same data sets, the overall computational time of the tools for alignment only (\ie, msalign, SpecAlign, and COW) were between tens of minutes to two hours (without peak fitting) depending on the data size. As discussed above, CAFA and ShapeFinder, the previous full suites available for nucleic acid CE quantification, required even more time (hours for the smaller data sets, extrapolated to days or weeks for the larger sets).  As shown in Figure~\ref{f:runtime}, the HiTRACE time breakdown is similar for all data sets, except for the 480-profile data set (P4-P6 CMCT), in which later stages are lengthened by increasing the number of bands in each profile (200 residues in the P4-P6 RNA, compared to under 100 residues for the other RNAs). {\color{red}We further used HiTRACE on data sets with longer RNAs (up to 400 nucleotides) and reverse transcribing from primers in the middle of a long RNA; the HiTRACE procedure was readily applied to these data sets (see supplement), and, encouragingly, the time demand remained linear with the number of bands.}

\section{Discussion}\label{s:discussion}
HiTRACE employs a series of automated techniques to control the high level of variability in parameters of CE systems and to resolve a key alignment bottleneck of modern nucleic acid structure mapping experiments. Several algorithmic advances are responsible for HiTRACE's accuracy and speed, including dynamic programming strategies that have not been been previously considered in the field. Quantitative comparisons on large experimental data sets demonstrate the utility of a linear time-axis transformation used in globally aligning profiles as well as the importance of a nonlinear alignment procedure for resolving further unavoidable variations in elution rates along a capillary. In addition, {\color{red} an interactive} band annotation {\color{red} interface} increases user convenience and provides accurate starting positions for the subsequent quantification step. These improvements have brought down the overall analysis time of data sets with tens of thousands of electrophoretic bands from days to minutes. {\color{red} The largest time-savings of the method are on experiments in which the same RNA sequence is probed under a variety of solution conditions, chemical modifiers, kinetic timepoints or mutations [see, e.g., ~\citep{daskaranicolasbaker2010,mitra2008high,weeksplos2009,weeks2010,kladwang2010,kladwangcordero2011}]. Now, the slow step in these and other experiments is interactive band annotation, which takes minutes (Fig.~\ref{f:runtime}). As more automated band assignment methods are developed (R.D., unpubl. results; personal comm., P. Pang, M. Elazar, J.S. Glenn), we plan to incorporate them into this interface.}


{\color{red}Although we designed HiTRACE primarily for RNA chemical structure mapping, the principles and premises that underlie HiTRACE are general and can easily be modified for use in other types of experimental assays. To enhance the adoption of this tool, we have created a stand-alone version of HiTRACE with a graphical user interface.} We are also making the source code freely available to encourage further innovation and incorporation of these algorithms into other laboratories' CE software suites. Beyond the data sets discussed herein, HiTRACE is continuously being used for other studies, totalling over 20,000 profiles (greater than 2 million bands) at the time of submission (unpubl. data, W.K., R.D.; see also http://rmdb.stanford.edu).  Given its accuracy, robustness and efficiency, we expect that HiTRACE will become a valuable tool for nucleic acid experimentalists entering a high-throughput era of structural analysis.

\section*{Acknowledgments}
The authors thank Dr. Alain Laederach at Wadsworth Center for providing CAFA sample data and members of the Das lab for comments on the manuscript and extensive testing.

\paragraph{Funding\textcolon}
This work was supported in part by the National Research Foundation of Korea funded by the Ministry of Education, Science and Technology (Grant No. 2010-0000407 and No. 2010-0000631 to SY) and in part by a Burroughs-Wellcome Foundation Career Award at the Scientific Interface (to RD for computational work).

\bibliographystyle{natbib}
\bibliography{hitrace}

\pagebreak
\section*{Supplement}
\appendix
\section{Additional details and examples}
\subsection{Step B.2 (inter-batch alignment)}
\begin{figure}[!h]
\centering
    \includegraphics[width=\linewidth]{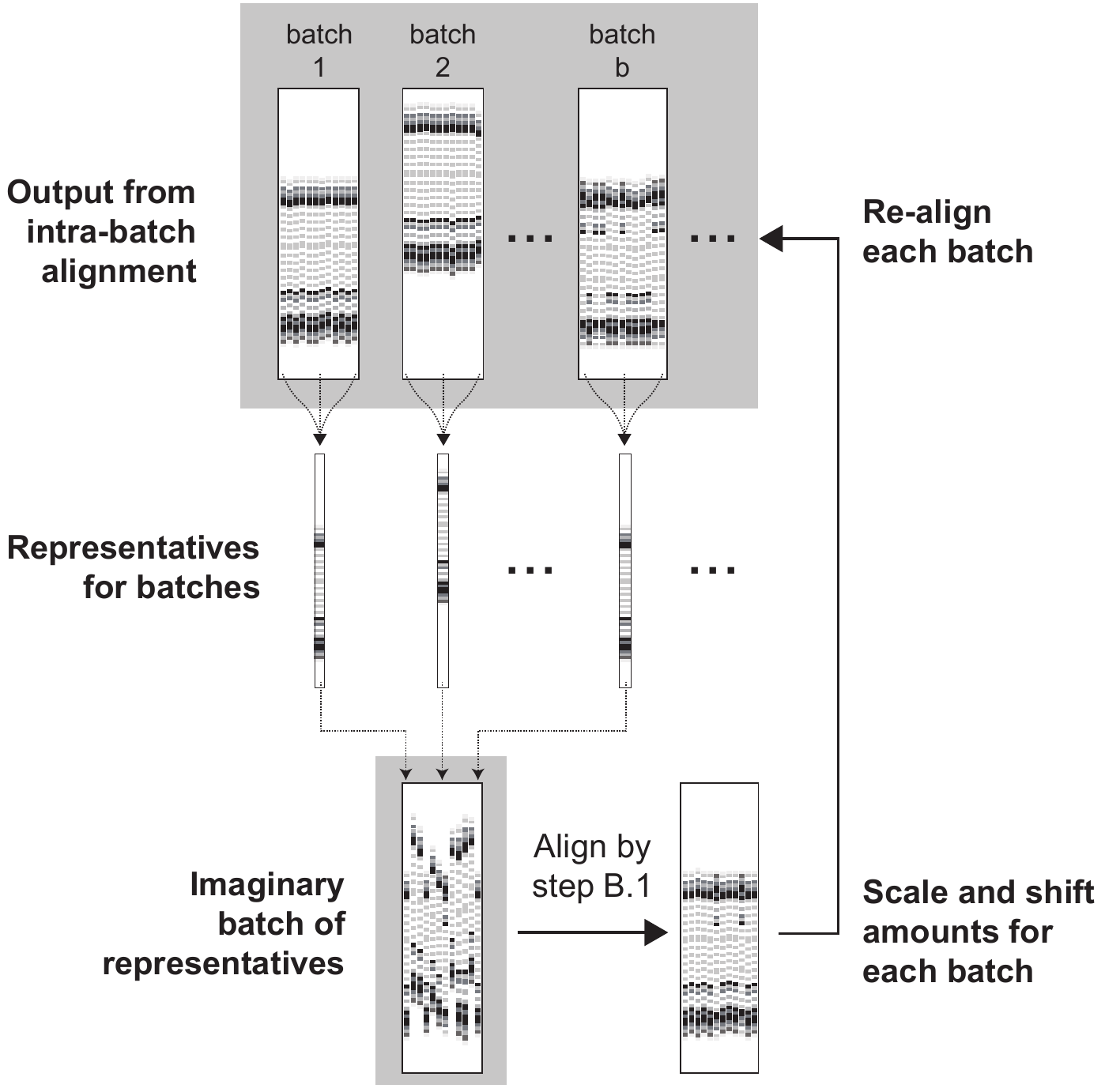}
\caption{Inter-batch alignment (step B.2). A representative profile is constructed for each batch that has been aligned in step B.1. All representatives are collected and then aligned by the intra-batch algorithm, as if these were from a single batch. The resulting scale and shift amounts for each batch are used for re-aligning the batch.
}
\label{f:fine-tuning}
\end{figure}

The alignment procedure in step B.1 can be considered as an \emph{intra-batch} step in that we separately align the fluorescence profiles in each batch without considering profiles in other batches. Due to variabilities between batches, performing only the intra-batch alignment above produces stratified alignment results, where a number of up-and-down `stairs' appear. To resolve this problem, we perform an additional \emph{inter-batch} alignment, as illustrated in Figure~\ref{f:fine-tuning}.

\subsection{Binarization-based alignment (step B.3; optional )}
This step can be optionally applied as the last of the correlation optimized linear alignment steps; because it did not improve precision assessed in cross-replicate correlation experiments, it is not performed in the default HiTRACE workflow. After inter-batch alignment, this step performs a peak detection on each profile and then binarize the profile so that the intensity at a peak position is set to 1 and the rest is set to 0. We then align the binarized profiles as before and use the resulting scale and shift information for re-aligning the original, non-binary profiles. This has the effect of low-pass filtering~\citep{oppenheim09} to suppress high-frequency noise components and aligns only peaks within each profile; it can give an improvement in alignment near the top of the data where multiple intense electrophoretic products overlap. For the peak detection process, we found that any reasonable peak detection method can be employed; we utilized the one described in~\citet{kim09}.

\subsection{Step C (nonlinear alignment)}
\begin{figure}
\centering
    \includegraphics[width=0.95\linewidth]{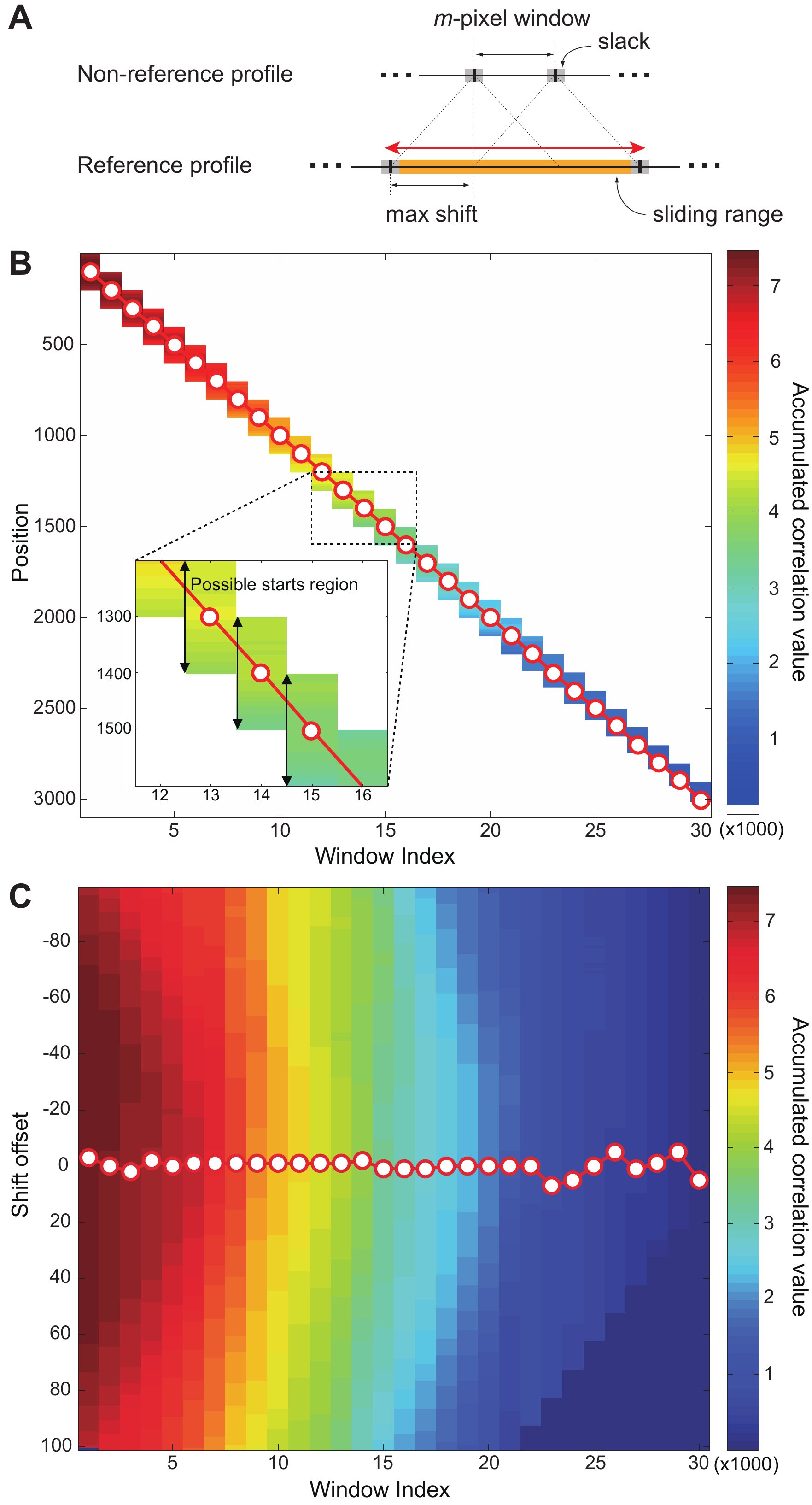}
\caption{Nonlinear alignment (step C). (A) We break the time axis of a non-reference profile into $m$-pixel windows and drift each window (in pixel units; each pixel is 0.1 seconds) within a predefined range over the reference profile to find the shift amount that maximizes the correlation of the window and the corresponding fragment of the reference profile. The `slack' is the amount by which we can extend or shrink each boundary of a window with respect to $m$, the default window size. The `max shift' is the largest difference possible between a window boundary in a non-reference profile and its corresponding boundary in the reference profile. The `sliding range' is the search region in the reference profile over which we compare a window from a non-reference profile. We find the optimal shift amount of each window by dynamic programming (DP). (B) Example of the score matrix for DP-based profile alignment. For each window, we determine its optimal shift offset using this matrix. The objective function is the total correlation coefficient value accumulated over all windows. Shown is the matrix for aligning the first and the sixteenth profile of the Medloop CMCT data (replicate 2) described in the result section. We set the window size $m$ to 100, producing 30 windows in total. The maximum amount of shifts allowed was 100, and the slack size used was 10. (C) A matrix to show the possible shift offsets of each window. Red circles indicate the optimal offsets determined by the backtracking procedure in (B).}
\label{f:nonlinear}
\end{figure}

The concept underlying the non-linear alignment step is depicted in Figure~\ref{f:nonlinear}A. Figure~\ref{f:nonlinear}B--C shows an example of determining the shift amount of each window edge for an actual fluorescence profile.

\subsection{Step D.2 (automated transfer of band annotation; optional)}
\begin{figure}
\centering
    \includegraphics[width=\linewidth]{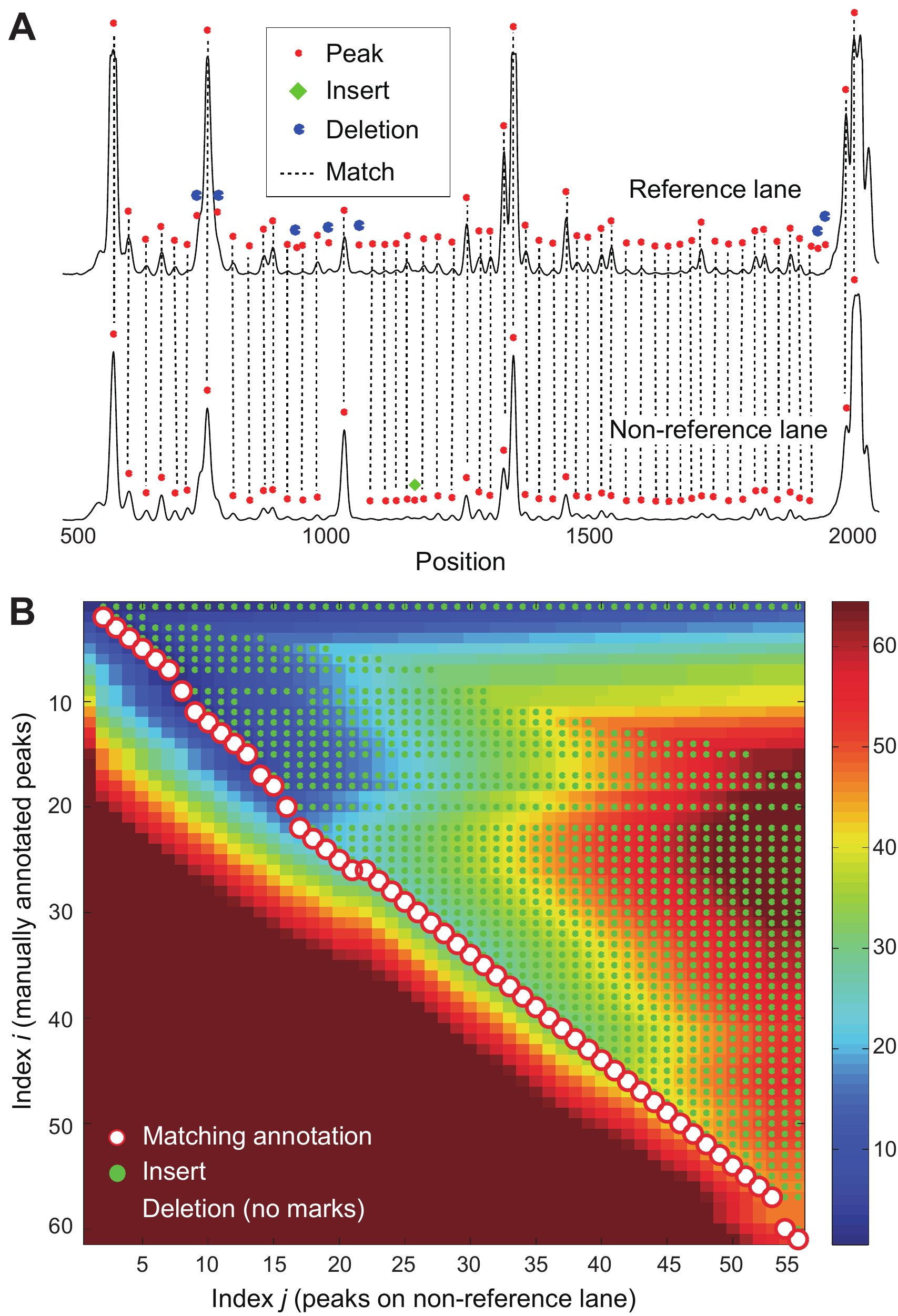}
\caption{Automated transfer of band annotation (step D.2; optional). (A) Some of the automatically found bands did not have a matching annotation in the reference profile. We name these extra bands \emph{inserts}. We do not find some sites in the manual annotation via automated peak fitting, and call these missing bands (\ie, false negatives) \emph{deletions}. Shown are the first and the 120-th profiles of the Medloop CMCT data (replicate 2) described in the result section. (B) Example of the score matrix $F(i,j)$ for dynamic-programming-based transfer of band annotation of the 120-th profile shown in (A). The optimal band assignments found by backtracking are shown using red circles.}
\label{f:auto-anno}
\end{figure}

Each band in a fluorescence profile corresponds to a position in the nucleic acid sequence. Given an annotation of one reference profile, HiTRACE can automatically annotate the other fluorescence profiles using a dynamic programming approach similar to the Needleman-Wunsch algorithm~\citep{needleman1970general}. This procedure was used before the development of nonlinear dynamic-programming-based alignment (step C); at that time, the final alignment of profiles was poorer in quality. With the current software including non-linear alignment, automated transfer of band annotation does not improve precision assessed in cross-replicate correlation experiments, so it is not performed in the default HiTRACE workflow. Nevertheless, we briefly summarize the annotation transfer algorithm here, as it can be carried out in HiTRACE (as the $guess\_all\_peaks$ script), and appears useful in a partially developed strategy for automated sequence assignment (R.D., unpub. results).

The procedure of transferring the annotation from the reference profile to all other profiles starts with identification of bands in each profile by a peak detector. Due to noise and imperfections in experiments and analysis, some of the automatically detected bands do not have a matching annotation (these are called \emph{inserts}), whereas some bands assigned in the manual annotation do not correspond to any automatically detected bands (\emph{deletions}). See Supplementary Figure 3 for an example. Transferring band annotations requires accurate identification of which bands are extraneous or missing in each non-reference profile, a task that we carry out through a dynamic programming strategy.


\begin{figure}
\centering
    \includegraphics[width=\linewidth]{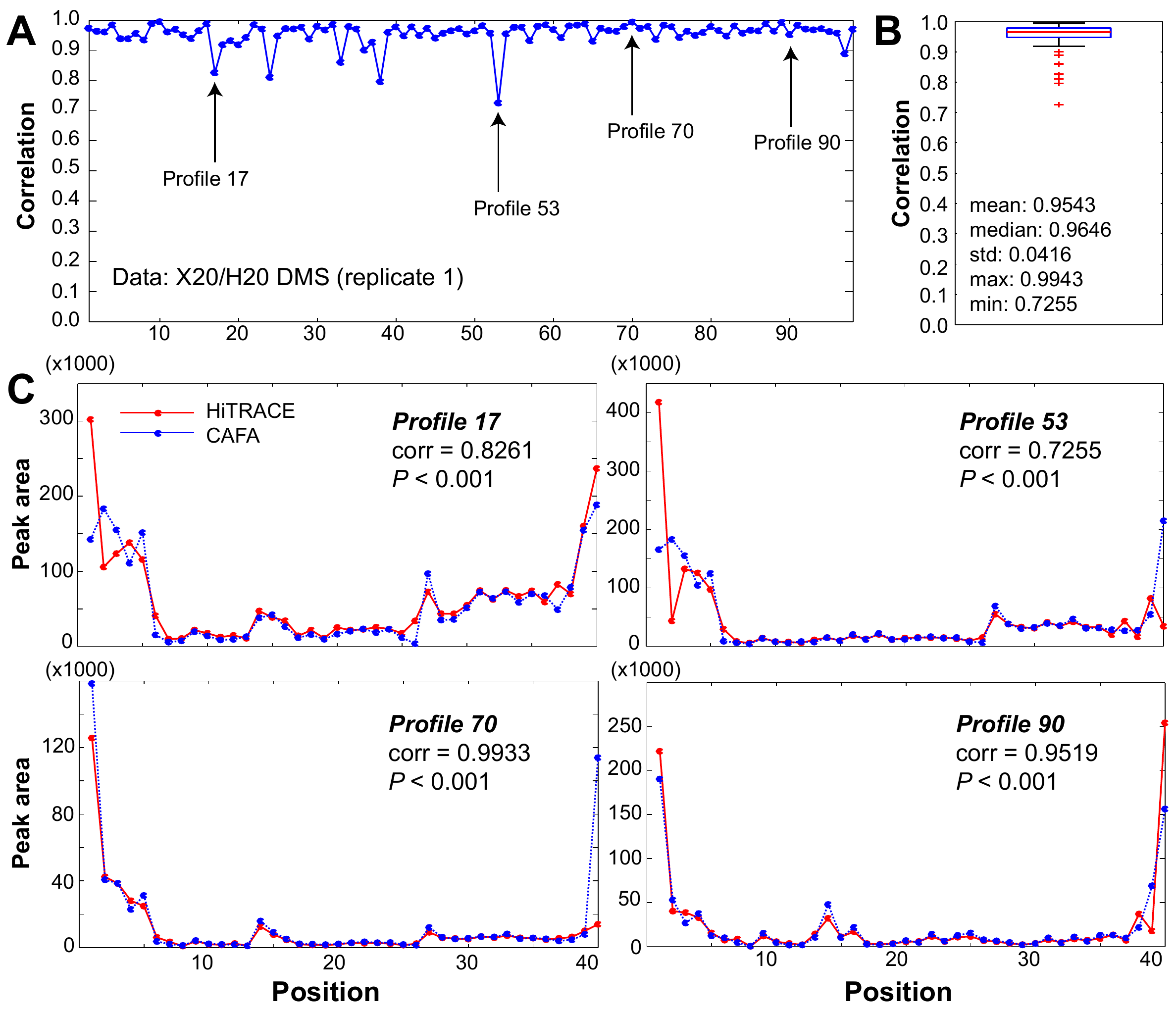}
\caption{Correlation between peak areas quantified by HiTRACE and CAFA~\citep{mitra2008high} for an additional data set. (A) Each point on the plot indicates the correlation coefficient between the peak areas HiTRACE and CAFA computed for an identical profile. The data used was X20/H20 DMS (replicate 1), which has 98 profiles. The time demands for quantification by HiTRACE and CAFA were approximately 4 minutes and 7 hours, respectively. (B) The distribution of the correlation coefficients. The average value was high (0.9543), suggesting that the results obtained from HiTRACE and CAFA are highly correlated for this data set. (C) More detailed plots for four arbitrary profiles. 
}
\label{f:corr-cafa-hitrace}
\end{figure}
Let sequence $R=\langle r_1, r_2,\ldots,r_i,\ldots \rangle$ denote the manually annotated band positions (in pixels) in the reference. Similarly, given a profile to be aligned to the reference profile, let sequence $A=\langle a_1, a_2,\ldots,a_j,\ldots \rangle$ denote the band locations. For dynamic programming, we build a score matrix $F$ indexed by $i$ and $j$ ($i$ for $R$ and $j$ for $A$), where the value $F(i,j)$ indicates the score of the best alignment between the prefix $\langle r_1,r_2,\ldots r_i\rangle$ of $R$ and the prefix $\langle a_1,a_2,\ldots, a_j\rangle$ of $A$. The matrix $F$ can be filled recursively by the following formula
\begin{equation}
F(i,j) = \min
    \begin{cases}
        F(i - 1, j - 1) + \mathrm{matchScore}(i,j)\\
        F(i - 1, j ) + \mathrm{deletionPenalty}(i)\\
        F(i, j - 1) + \mathrm{insertPenalty}
    \end{cases}
\end{equation}
after a trivial initialization process. Backtracking on the matrix $F$ reveals the optimal assignment of automatically found bands to manual annotations (Figure~\ref{f:auto-anno}). For any deletions, we estimated band locations missing in the non-reference profile based on linear interpolation between the nearest bands that match in the reference and non-reference profile.

We considered a few factors to define $\mathrm{matchScore}(i,j)$; these functional forms and presented parameter settings were defined based on empirical results on one large-scale data set (P4-P6 SHAPE; see Table~1 in the main article); the other data sets present independent tests of these parameters. The match penalty is a weighted sum of four factors:
\begin{equation}
\mathrm{matchScore}(i,j) = \sum_{k=1}^4 w_k\cdot\mathrm{matchScore}_k(i,j)
\end{equation}
where $w_k$ is the weight of factor $k$. First, we let the match penalty proportional to the distance between $r_i$ and $a_j$, penalizing distant matches. The first component is thus given by
\begin{equation}
\mathrm{matchScore}_1(i,j) =\left|\frac{r_i-a_j}{d_R} \right|^2
\end{equation}
where $d_R$ is the average distance between two adjacent reference peaks in sequence $R$.
Second, we consider the degree of peak-to-peak separation as follows:
\begin{equation}
\mathrm{matchScore}_2(i,j)=\left|\frac{ (r_i-r_{i^*})-(a_j-a_{j^*})}{d_R\cdot(i-i^*)}\right|^2
\end{equation}
where $i^*$ and $j^*$ represent the position of the previous matching pair.
Third, we consider the difference between the intensity of $r_i$ and $a_j$ relative to the previous matching location:
\begin{equation}
\mathrm{matchScore}_3(i,j) = \left | \log\frac{I(r_i)}{I(r_{i^*})} - \log\frac{I(a_j)}{I(a_{j^*})} \right |
\end{equation}
where $I(\cdot)$ represents the profile intensity.
Lastly, we reward band assignments to points of greater intensity up to a point by defining the last component:
\begin{equation}
\mathrm{matchScore}_4(i,j) = 1 - \min \left\{ \frac{I(a_j)}{I_R}, 1 \right\}
\end{equation}
where $I_R$ is the median intensity of the reference profile. The weights used are $w_1=1$, $w_2=4$, $w_3=0.25$ and $w_4=2$.

The $\mathrm{insertPenalty}$ was set to 1.5. To determine $\mathrm{deletionPenalty}(i)$, we used an expectation-maximization (EM) approach~\citep{bishop06}. After a first run with an initial constant value (4.5) for deletion penalty, we carried out a second run with deletion penalties inversely proportional to deletion frequencies at each peak position $i$ seen in the first run.

\bibliographystyle{natbib}
\bibliography{hitrace}

\begin{thebibliography}{}

\bibitem[Bishop(2006)Bishop]{bishop06}
Bishop, C.~M. (2006).
\newblock {\em Pattern recognition and machine learning\/}.
\newblock Springer, New York.

\bibitem[Bylund {\em et~al.}(2002)Bylund, Danielsson, Malmquist, and
  Markides]{bylund2002chromatographic}
Bylund, D., Danielsson, R., Malmquist, G., and Markides, K. (2002).
\newblock {Chromatographic alignment by warping and dynamic programming as a
  pre-processing tool for PARAFAC modelling of liquid chromatography-mass
  spectrometry data}.
\newblock {\em Journal of Chromatography A\/}, {\bf 961}(2), 237--244.

\bibitem[Cormen {\em et~al.}(2009)Cormen, Leiserson, Rivest, and
  Stein]{cormen2009introduction}
Cormen, T., Leiserson, C., Rivest, R., and Stein, C. (2009).
\newblock {\em {Introduction to algorithms}\/}.
\newblock The MIT press, Cambridge, Massachusetts, 3rd edition.

\bibitem[Cover and Thomas(2006)Cover and Thomas]{cover2006elements}
Cover, T. and Thomas, J. (2006).
\newblock {\em {Elements of information theory}\/}.
\newblock John Wiley and Sons, Hoboken, New Jersey, 2nd edition.

\bibitem[Das {\em et~al.}(2005)Das, Laederach, Pearlman, Herschlag, and
  Altman]{das2005safa}
Das, R., Laederach, A., Pearlman, S., Herschlag, D., and Altman, R. (2005).
\newblock {SAFA}: semi-automated footprinting analysis software for
  high-throughput quantification of nucleic acid footprinting experiments.
\newblock {\em RNA\/}, {\bf 11}(3), 344--354.

\bibitem[Das {\em et~al.}(2010)Das, Karanicolas, and
  Baker]{daskaranicolasbaker2010}
Das, R., Karanicolas, J., and Baker, D. (2010).
\newblock Atomic accuracy in predicting and designing noncanonical {RNA}
  structure.
\newblock {\em Nature Methods\/}, {\bf 7}(4), 291--294.

\bibitem[Deigan {\em et~al.}(2009)Deigan, Li, Mathews, and
  Weeks]{deigan2009accurate}
Deigan, K., Li, T., Mathews, D., and Weeks, K. (2009).
\newblock {Accurate {SHAPE}-directed {RNA} structure determination}.
\newblock {\em Proceedings of the National Academy of Sciences\/}, {\bf
  106}(1), 97.

\bibitem[Ewing and Green(1998)Ewing and Green]{phred2}
Ewing, B. and Green, P. (1998).
\newblock Base-calling of automated sequencer traces using {Phred.} {II.} error
  probabilties.
\newblock {\em Genome Research\/}, {\bf 8}(3), 186--194.

\bibitem[Ewing {\em et~al.}(1998)Ewing, Hillier, Wendl, and Green]{phred1}
Ewing, B., Hillier, L., Wendl, M.~C., and Green, P. (1998).
\newblock Base-calling of automated sequencer traces {Using Phred.} {I.}
  accuracy assessment.
\newblock {\em Genome Research\/}, {\bf 8}(3), 175--185.

\bibitem[Kay(1993)Kay]{kay1993fundamentals}
Kay, S. (1993).
\newblock {\em {Fundamentals of statistical signal processing: estimation
  theory}\/}.
\newblock Prentice Hall, Upper Saddle River, New Jersey.

\bibitem[Kazmi {\em et~al.}(2006)Kazmi, Ghosh, Shin, Hill, and
  Grant]{kazmi2006alignment}
Kazmi, S., Ghosh, S., Shin, D., Hill, D., and Grant, D. (2006).
\newblock {Alignment of high resolution mass spectra: Development of a
  heuristic approach for metabolomics}.
\newblock {\em Metabolomics\/}, {\bf 2}(2), 75--83.

\bibitem[Kim {\em et~al.}(2009)Kim, Yu, Shim, Kim, Min, Chung, Das, and
  Yoon]{kim09}
Kim, J., Yu, S., Shim, B., Kim, H., Min, H., Chung, E.~Y., Das, R., and Yoon,
  S. (2009).
\newblock A robust peak detection method for {RNA} structure inference by
  high-throughput contact mapping.
\newblock {\em Bioinformatics\/}, {\bf 25}(9), 1137--44.

\bibitem[Kladwang and Das(2010)Kladwang and Das]{kladwang2010}
Kladwang, W. and Das, R. (2010).
\newblock A mutate-and-map strategy for inferring base pairs in structured
  nucleic acids: proof of concept on a {DNA/RNA} helix.
\newblock {\em Biochemistry\/}, {\bf 49}(35), 7414--7416.

\bibitem[Kladwang {\em et~al.}(2011)Kladwang, Cordero, and
  Das]{kladwangcordero2011}
Kladwang, W., Cordero, P., and Das, R. (2011).
\newblock {A mutate-and-map strategy accurately infers the base pairs of an
  35-nucleotide model RNA}.
\newblock {\em RNA\/}, {\bf 17}, 522--534.

\bibitem[Laederach {\em et~al.}(2008)Laederach, Das, Vicens, Pearlman,
  Brenowitz, Herschlag, and Altman]{laederach2008semiautomated}
Laederach, A., Das, R., Vicens, Q., Pearlman, S., Brenowitz, M., Herschlag, D.,
  and Altman, R. (2008).
\newblock {Semiautomated and rapid quantification of nucleic acid footprinting
  and structure mapping experiments}.
\newblock {\em Nature Protocols\/}, {\bf 3}(9), 1395--1401.

\bibitem[Levenberg(1944)Levenberg]{levenberg1944method}
Levenberg, K. (1944).
\newblock {A method for the solution of certain nonlinear problems in least
  squares}.
\newblock {\em Quart. Appl. Math\/}, {\bf 2}(2), 164--168.

\bibitem[Marquardt(1963)Marquardt]{marquardt1963algorithm}
Marquardt, D. (1963).
\newblock {An algorithm for least-squares estimation of nonlinear parameters}.
\newblock {\em Journal of the Society for Industrial and Applied
  Mathematics\/}, {\bf 11}(2), 431--441.

\bibitem[Merino {\em et~al.}(2005)Merino, Wilkinson, Coughlan, and
  Weeks]{wilkinson2005}
Merino, E., Wilkinson, K., Coughlan, J., and Weeks, K. (2005).
\newblock {Advances in RNA structure analysis by chemical probing}.
\newblock {\em J. Am. Chem. Soc.}, {\bf 127}, 4223--4231.

\bibitem[Mitra {\em et~al.}(2008)Mitra, Shcherbakova, Altman, Brenowitz, and
  Laederach]{mitra2008high}
Mitra, S., Shcherbakova, I., Altman, R., Brenowitz, M., and Laederach, A.
  (2008).
\newblock {High-throughput single-nucleotide structural mapping by capillary
  automated footprinting analysis}.
\newblock {\em Nucleic Acids Research\/}, {\bf 36}(11), e63.

\bibitem[Needleman and Wunsch(1970)Needleman and Wunsch]{needleman1970general}
Needleman, S. and Wunsch, C. (1970).
\newblock {A general method applicable to the search for similarities in the
  amino acid sequence of two proteins}.
\newblock {\em Journal of molecular biology\/}, {\bf 48}(3), 443--453.

\bibitem[Nielsen {\em et~al.}(1998)Nielsen, Carstensen, and
  Smedsgaard]{nielsen1998aligning}
Nielsen, N., Carstensen, J., and Smedsgaard, J. (1998).
\newblock {Aligning of single and multiple wavelength chromatographic profiles
  for chemometric data analysis using correlation optimised warping}.
\newblock {\em Journal of Chromatography A\/}, {\bf 805}(1-2), 17--35.

\bibitem[Oppenheim and Schafer(2009)Oppenheim and Schafer]{oppenheim09}
Oppenheim, A.~V. and Schafer, R.~W. (2009).
\newblock {\em Discrete-time signal processing\/}.
\newblock Prentice Hall, Upper Saddle River, New Jersey, 3rd edition.

\bibitem[Peattie and Gilbert(1980)Peattie and Gilbert]{peattie1980chemical}
Peattie, D. and Gilbert, W. (1980).
\newblock {Chemical probes for higher-order structure in RNA}.
\newblock {\em Proceedings of the National Academy of Sciences of the United
  States of America\/}, {\bf 77}(8), 4679.

\bibitem[Pravdova {\em et~al.}(2002)Pravdova, Walczak, and
  Massart]{pravdova2002comparison}
Pravdova, V., Walczak, B., and Massart, D. (2002).
\newblock {A comparison of two algorithms for warping of analytical signals}.
\newblock {\em Analytica Chimica Acta\/}, {\bf 456}(1), 77--92.

\bibitem[Robinson {\em et~al.}(2007)Robinson, De~Souza, Keen, Saunders,
  McConville, Speed, and Liki{\'c}]{robinson2007dynamic}
Robinson, M., De~Souza, D., Keen, W., Saunders, E., McConville, M., Speed, T.,
  and Liki{\'c}, V. (2007).
\newblock {A dynamic programming approach for the alignment of signal peaks in
  multiple gas chromatography-mass spectrometry experiments}.
\newblock {\em BMC bioinformatics\/}, {\bf 8}(1), 419.

\bibitem[Ruiz-Martinez {\em et~al.}(1993)Ruiz-Martinez, Berka, Belenkii, Foret,
  Miller, and Karger]{ruiz1993dna}
Ruiz-Martinez, M., Berka, J., Belenkii, A., Foret, F., Miller, A., and Karger,
  B. (1993).
\newblock {DNA sequencing by capillary electrophoresis with replaceable linear
  polyacrylamide and laser-induced fluorescence detection}.
\newblock {\em Analytical chemistry\/}, {\bf 65}(20), 2851--2858.

\bibitem[Tijerina {\em et~al.}(2007)Tijerina, Mohr, and
  Russell]{tijerina2007dms}
Tijerina, P., Mohr, S., and Russell, R. (2007).
\newblock {DMS footprinting of structured RNAs and RNA--protein complexes}.
\newblock {\em Nature Protocols\/}, {\bf 2}(10), 2608--2623.

\bibitem[Tomasi {\em et~al.}(2004)Tomasi, van~den Berg, and
  Andersson]{tomasi2004correlation}
Tomasi, G., van~den Berg, F., and Andersson, C. (2004).
\newblock {Correlation optimized warping and dynamic time warping as
  preprocessing methods for chromatographic data}.
\newblock {\em Journal of Chemometrics\/}, {\bf 18}(5), 231--241.

\bibitem[Vasa {\em et~al.}(2008)Vasa, Guex, Wilkinson, Weeks, and
  Giddings]{vasa2008shapefinder}
Vasa, S., Guex, N., Wilkinson, K., Weeks, K., and Giddings, M. (2008).
\newblock {ShapeFinder: A software system for high-throughput quantitative
  analysis of nucleic acid reactivity information resolved by capillary
  electrophoresis}.
\newblock {\em RNA\/}, {\bf 14}(10), 1979--1990.

\bibitem[Walczak {\em et~al.}(1996)Walczak, Westhof, Carbon, and Krol]{walczak}
Walczak, R., Westhof, E., Carbon, P., and Krol, A. (1996).
\newblock {A novel RNA structural motif in the selenocysteine insertion element
  of eukaryotic selenoprotein mRNAs}.
\newblock {\em Current opinion in structural biology\/}, {\bf 2}, 367--379.

\bibitem[Watts {\em et~al.}(2009)Watts, Dang, Gorelick, Leonard, Bess,
  Swanstrom, Burch, and Weeks]{weeksnature2009}
Watts, J.~M., Dang, K.~K., Gorelick, R.~J., Leonard, C.~W., Bess, J.~W.,
  Swanstrom, R., Burch, C.~L., and Weeks, K.~M. (2009).
\newblock Architecture and secondary structure of an entire hiv-1 rna genome.
\newblock {\em Nature\/}, {\bf 460}(7256), 711--716.

\bibitem[Weeks(2010)Weeks]{weeks2010}
Weeks, K. (2010).
\newblock {Advances in RNA structure analysis by chemical probing}.
\newblock {\em Current opinion in structural biology\/}, {\bf 20}, 295--304.

\bibitem[Wilkinson {\em et~al.}(2008)Wilkinson, Gorelick, Vasa, Guex, Rein,
  Mathews, Giddings, and Weeks]{weeksplos2009}
Wilkinson, K.~A., Gorelick, R.~J., Vasa, S.~M., Guex, N., Rein, A., Mathews,
  D.~H., Giddings, M.~C., and Weeks, K.~M. (2008).
\newblock High-throughput {SHAPE} analysis reveals structures in {HIV-1}
  genomic {RNA} strongly conserved across distinct biological states.
\newblock {\em PLoS Biol\/}, {\bf 6}(4), e96+.

\bibitem[Wong {\em et~al.}(2005)Wong, Cagney, and
  Cartwright]{wong2005specalign}
Wong, J., Cagney, G., and Cartwright, H. (2005).
\newblock {SpecAlign--processing and alignment of mass spectra datasets}.
\newblock {\em Bioinformatics\/}, {\bf 21}(9), 2088--2090.

\bibitem[Woolley and Mathies(1995)Woolley and Mathies]{woolley1995ultra}
Woolley, A. and Mathies, R. (1995).
\newblock {Ultra-high-speed DNA sequencing using capillary electrophoresis
  chips}.
\newblock {\em Analytical chemistry\/}, {\bf 67}(20), 3676--3680.

\bibitem[Xi and Rocke(2008)Xi and Rocke]{xi2008baseline}
Xi, Y. and Rocke, D. (2008).
\newblock {Baseline correction for NMR spectroscopic metabolomics data
  analysis}.
\newblock {\em BMC bioinformatics\/}, {\bf 9}(1), 324.

\end{thebibliography}

\newpage

\begin{figure}
\centering
    \includegraphics[width=0.75\linewidth]{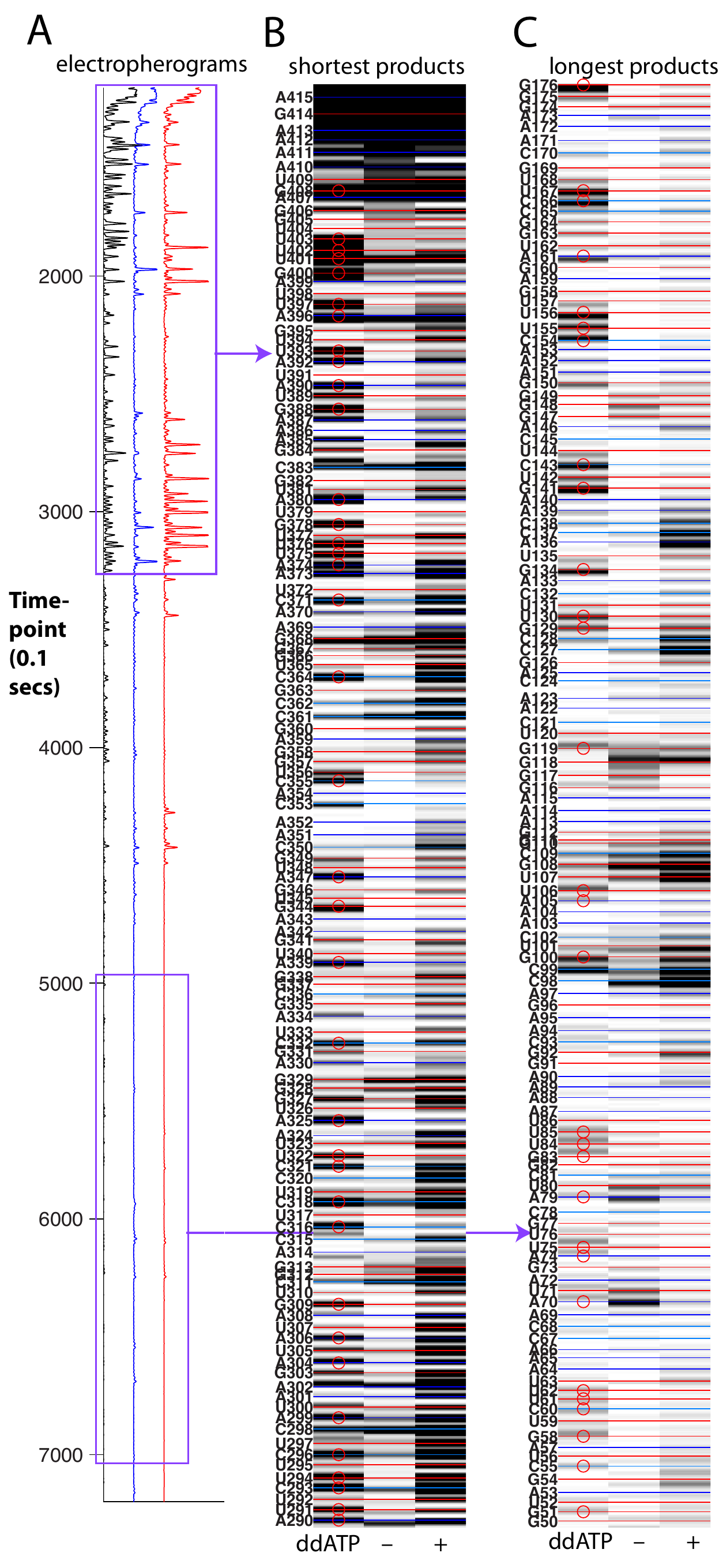}
\caption{Rapid  HiTRACE annotation for longer RNAs. (A) Capillary electropherograms for an experiment probing the ``unfolded'' L-21 ScaI ribozyme in 50 mM Na-HEPES, pH 8.0 (W.K., R.D., unpub. results). The fluorescence profiles (arbitrary units) are, from left to right, ddATP sequencing ladder, control reaction with no chemical modifier, and experiment with the NMIA reagent (for SHAPE acylation). (B) View in HiTRACE near the `top' of the data; note guidemark symbols in ddATP ladder. (C) View in HiTRACE near the `bottom' of the data.}
\label{f:L21-400nt}
\end{figure}

\begin{figure}
\centering
    \includegraphics[width=0.6\linewidth]{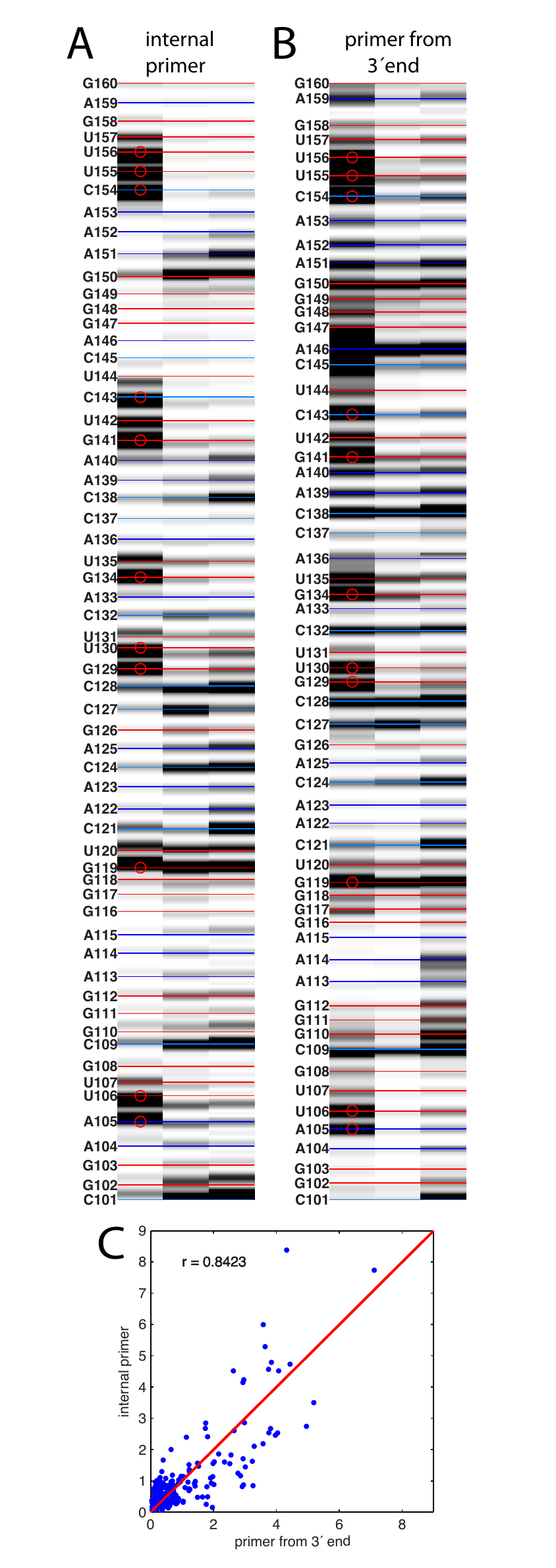}
\caption{Rapid HiTRACE annotation for reverse transcription from primers internal to an RNA sequnce. Data were collected for the P4-P6 RNA with primers to (A) the middle of this RNA's sequence (position 170) and to (B) the RNA's 3$^\prime$ end (position 270). For both sets, the fluorescence data are, from left to right, ddATP sequencing ladder, control reaction with no chemical modifier, and experiment with the NMIA reagent (for SHAPE acylation). (C) Correlation between the independent data sets.}
\label{f:P4P6-internal-prim}
\end{figure}
\newpage

\begin{figure*}
\centering
    \includegraphics[width=0.95\linewidth]{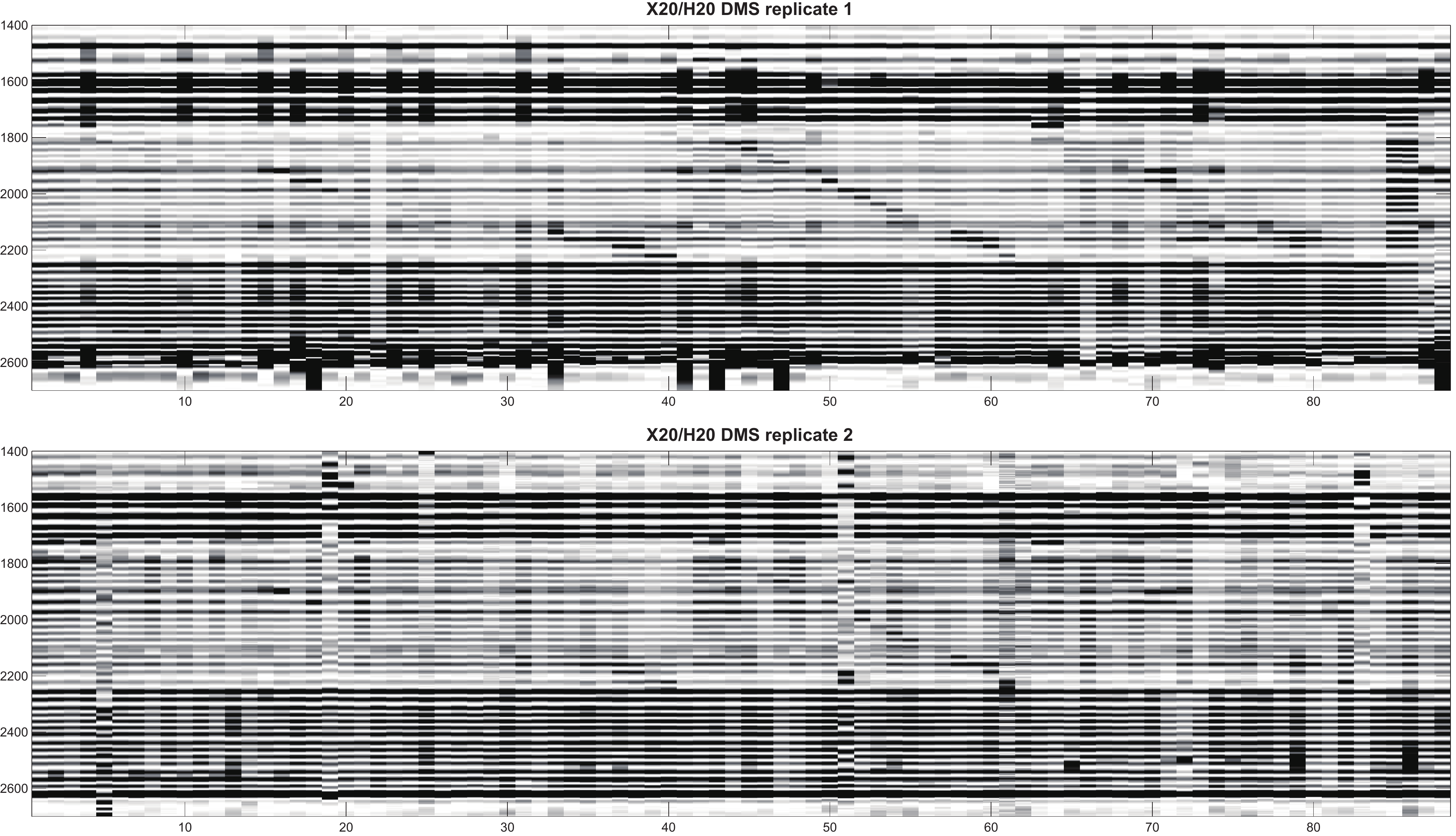}
\caption{Profiles from experimental replicates of X20/H20 DMS data after automated alignment refinement by dynamic-programming-based nonlinear adjustments ($x$-axis: profile, $y$-axis: band position).}
\label{f:x20-dms}
\end{figure*}

\begin{figure*}
\centering
    \includegraphics[width=0.95\linewidth]{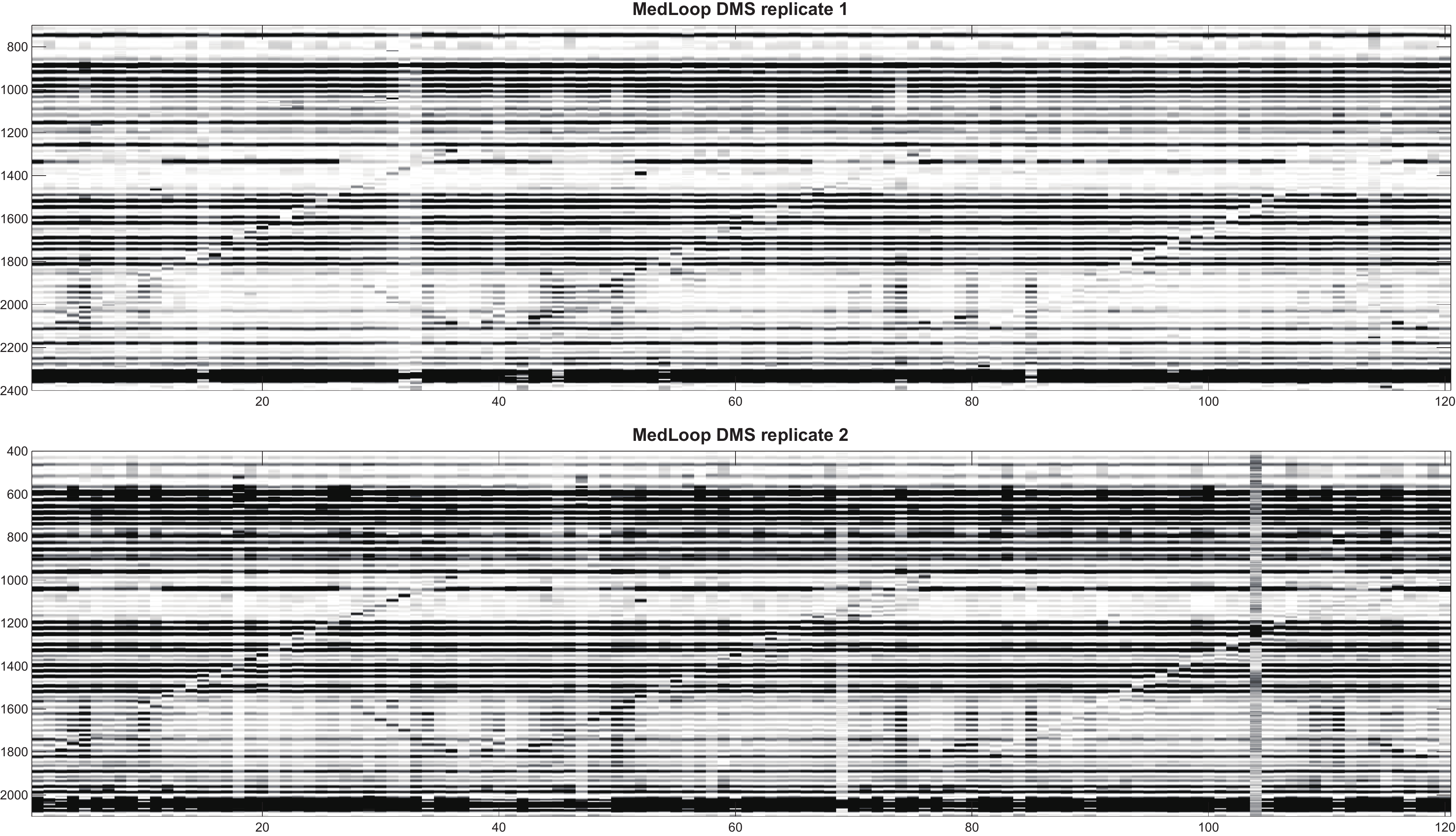}
\caption{Profiles from experimental replicates of Medloop DMS data.}
\label{f:medloop-dms}
\end{figure*}

\begin{figure*}
\centering
    \includegraphics[width=0.95\linewidth]{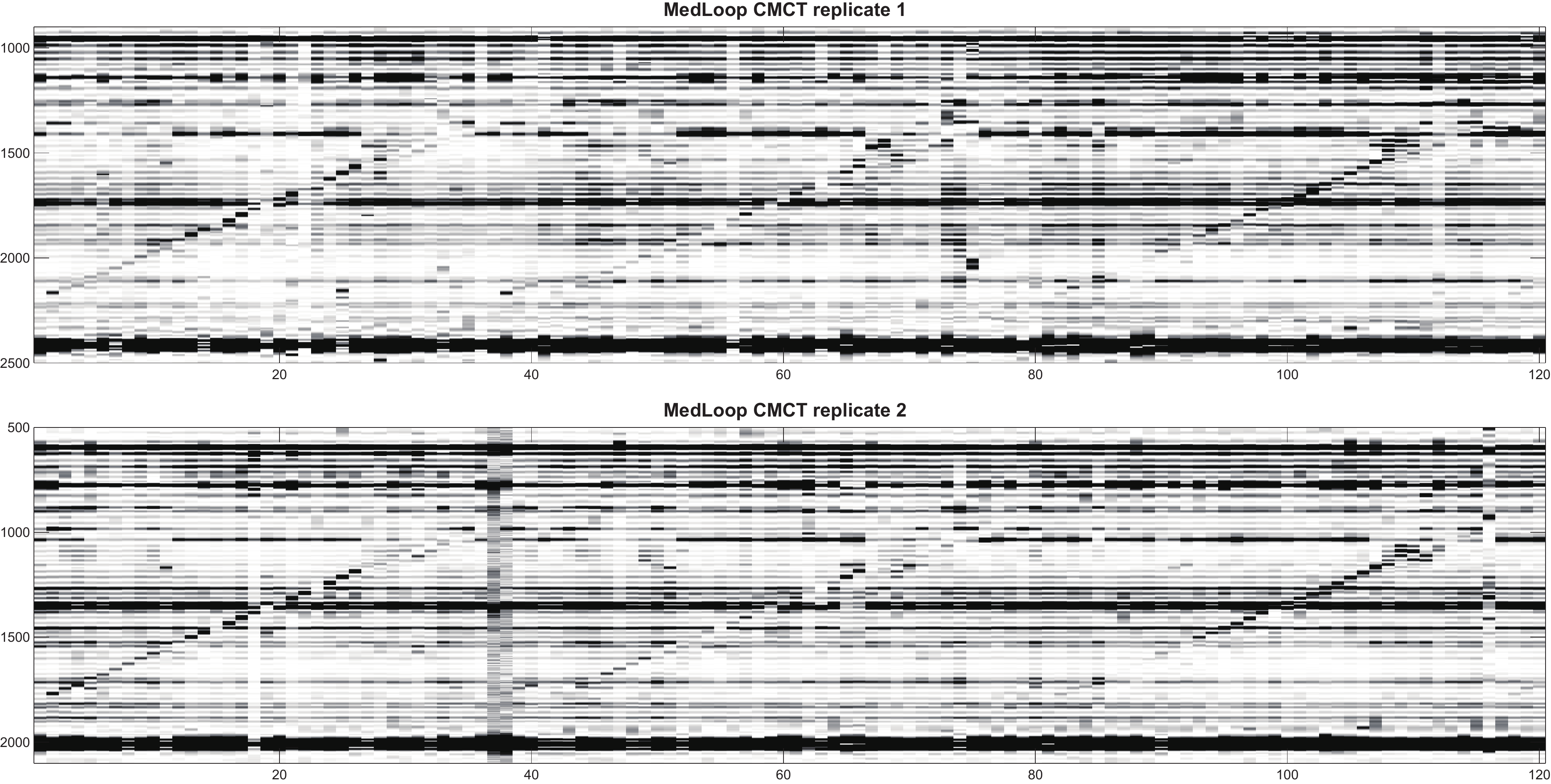}
\caption{Profiles from experimental replicates of Medloop CMCT data.}
\label{f:medloop-cmct}
\end{figure*}

\begin{figure*}
\centering
    \includegraphics[width=0.95\linewidth]{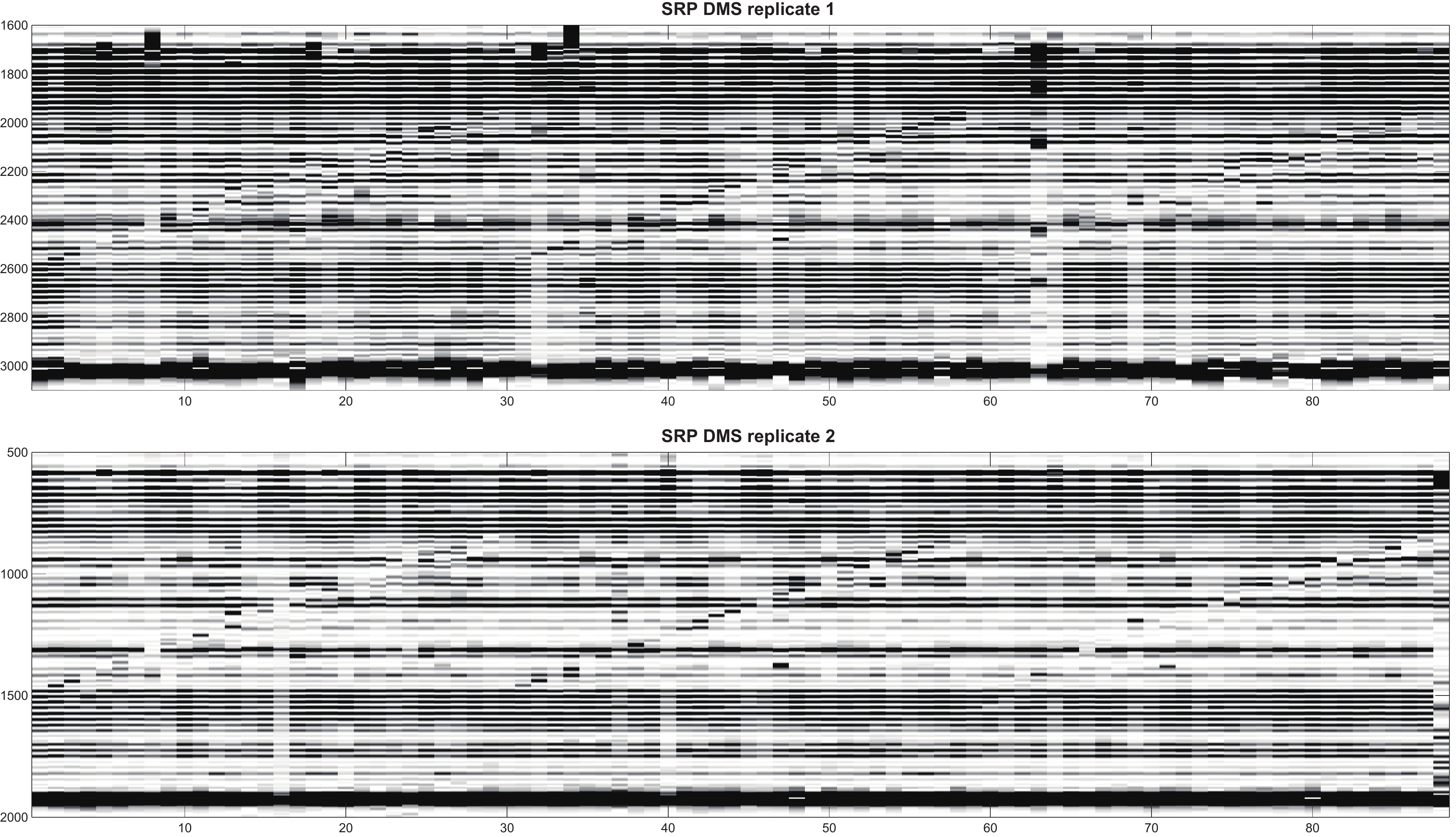}
\caption{Profiles from experimental replicates of SRP DMS data.}
\label{f:srp-dms}
\end{figure*}

\begin{figure*}
\centering
    \includegraphics[width=0.95\linewidth]{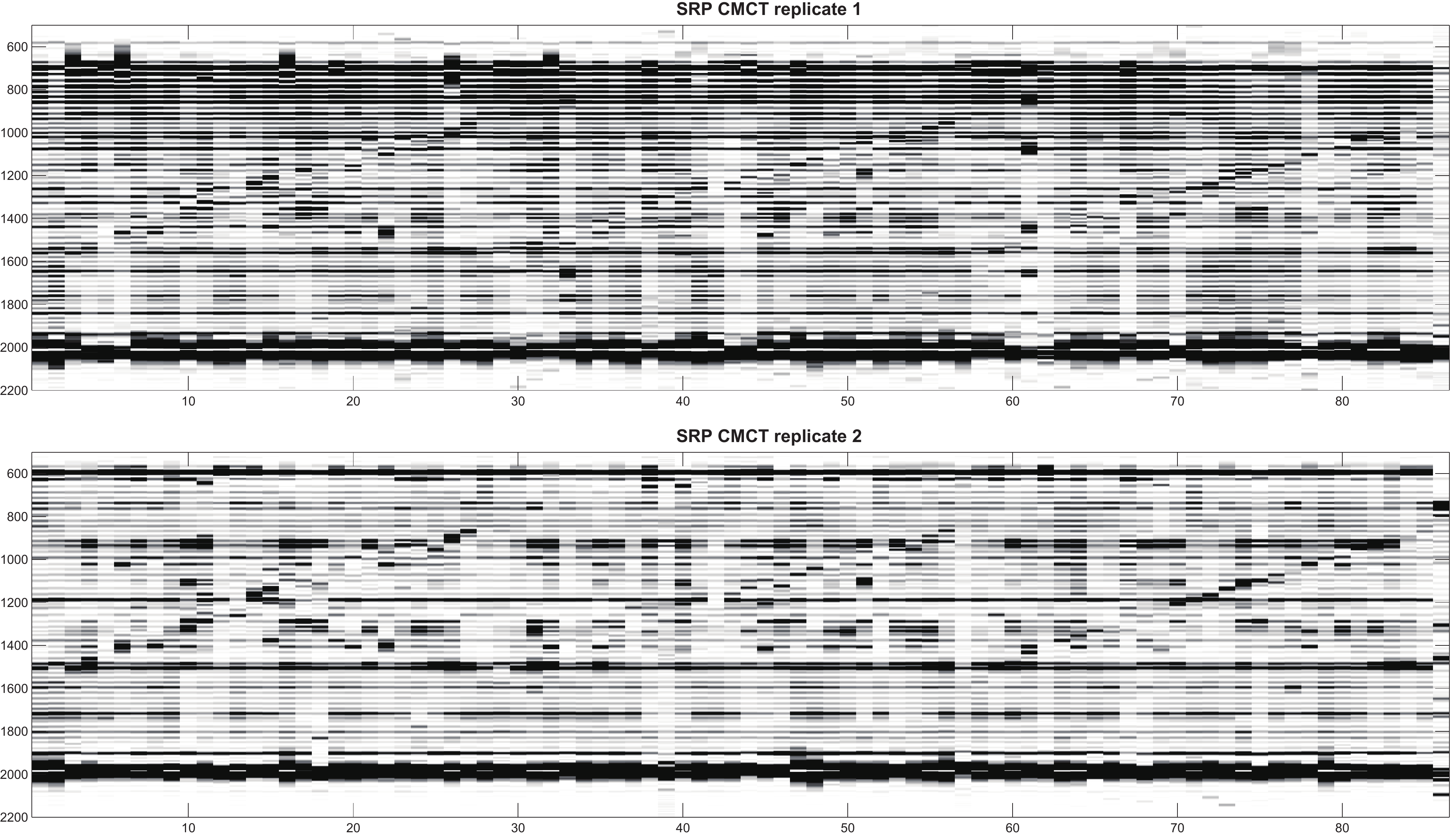}
\caption{Profiles from experimental replicates of SRP CMCT data.}
\label{f:srp-cmct}
\end{figure*}

\end{document}